\documentclass{article}

\usepackage{PRIMEarxiv}

\usepackage[utf8]{inputenc} 
\usepackage[T1]{fontenc}    
\usepackage{hyperref}       
\usepackage{url}            
\usepackage{booktabs}       
\usepackage{amsmath}
\usepackage{amsthm}
\usepackage{amssymb}
\usepackage{amsfonts}
\usepackage{mathtools}
\usepackage{amsfonts}       
\usepackage{nicefrac}       
\usepackage{microtype}      
\usepackage{fancyhdr}       
\usepackage{graphicx}       
\graphicspath{{media/}}     
\usepackage{times}
\usepackage{subfig}
\usepackage{xcolor}
\usepackage{array}
\usepackage{cite}
\usepackage{multirow}
\usepackage{adjustbox}
\usepackage[scr=dutchcal]{mathalfa}
\usepackage[font=small,labelfont=bf]{caption}
\usepackage{tikz-network}

\DeclareMathOperator{\sign}{sign}
\DeclareMathOperator{\diag}{diag}
\DeclareMathOperator{\tr}{tr}

\newtheorem{theorem}{Theorem}
\newtheorem{lemma}{Lemma}

\newtheorem{assumption}{Assumption}

\newtheorem{remark}{Remark}

\title{Distributed Adaptive Control of Disturbed Interconnected Systems with High-Order Tuners\thanks{This material is based upon work supported in by grants ONR N00014-21-1-2431, NSF 2121121, NSF 2208182, the U.S. Department of Homeland Security under Grant Award Number 22STESE00001-03-02, and by the Army Research Laboratory under Cooperative Agreement Number W911NF-22-2-0001. The views and conclusions contained in this document are solely those of the authors and should not be interpreted as representing the official policies, either expressed or implied, of the U.S. Department of Homeland Security, the Army Research Office, or the U.S. Government.} 
}

\author{
  Moh. Kamalul Wafi \\
  Department of Electrical and Computer Engineering \\
  Northeastern University \\
  Boston \\
  \texttt{wafi.m@northeastern.edu} \\
   \And
  Milad Siami \\
  Department of Electrical and Computer Engineering \\
  Northeastern University \\
  Boston \\
  \texttt{m.siami@northeastern.edu} \\
}

\begin{document}
\maketitle

\begin{abstract}
This paper addresses the challenge of network synchronization under limited communication, involving heterogeneous agents with different dynamics and various network topologies, to achieve consensus. We investigate the distributed adaptive control for interconnected unknown linear subsystems with a leader and followers, in the presence of input-output disturbance.  We enhance the communication within multi-agent systems to achieve consensus under the leadership's guidance. While the measured variable is similar among the followers, the incoming measurements are weighted and constructed based on their proximity to the leader. 
We also explore the convergence rates across various balanced topologies (Star-like, Cyclic-like, Path, Random), featuring different numbers of agents, using three distributed algorithms, ranging from first- to high-order tuners to effectively address time-varying regressors.
The mathematical foundation is rigorously presented from the network designs of the unknown agents following a leader, to the distributed methods. Moreover, we conduct several numerical simulations across various networks, agents and tuners to evaluate the effects of sparsity in the interaction between subsystems using the $L_2-$norm and $L_\infty-$norm. 
Some networks exhibit a trend where an increasing number of agents results in smaller errors, although this is not universally the case. Additionally, patterns observed at initial times may not reliably predict overall performance across different networks. Finally, we demonstrate that the proposed modified high-order tuner outperforms its counterparts, and we provide related insights along with our conclusions.
\end{abstract}
\allowdisplaybreaks

\section{Introduction}\label{S1}
Multi-agent systems (MAS), spanning areas from robotics, including unmanned ground \cite{R1}, aerial \cite{R2}, and underwater vehicles \cite{R3}, to large-scale societal dynamics \cite{R4}, have attracted considerable interest.
The scope of challenges these systems face extends from internal issues like achieving consensus among agents for coordinated control and stability, to external threats such as disturbances, environmental uncertainties, or attacks
\cite{R5}. Furthermore, 
the interconnected nature of networked systems necessitates insights from graph theory. This is underscored by \cite{R6}, which examines the limits and trade-offs in networks facing stochastic disturbances, and by \cite{R7}, which explores how denser networks (with more links) affect the number of agents.

In this paper, we explore distributed adaptive control as a foundational element of MAS, sharing similarities with distributed Model Reference Adaptive Control (MRAC). This topic spans from theoretical frameworks aimed at achieving consensus \cite{R8,R9} in complex networks to practical applications in large-scale systems \cite{R10,R11}. The distributed adaptive control in this paper is adaptable to various agent dynamics, influencing diverse control laws among agents, as closely discussed in \cite{R12,R13}. Moreover, specific studies have proposed solutions for nonlinear MAS and neural network-based challenges \cite{R14,R15}. Furthermore, drawing inspiration from recent advancements in distributed optimization \cite{R16}, our study incorporates high-order tuners as adaptive laws to update the gains.


A newly developed algorithm for high-order tuners has been introduced to optimize convex loss functions with time-varying regressors in identification problems. This algorithm leverages Nesterov’s method principles, ensuring that parameter estimations stay within predefined bounds when confronted with time-varying regressors \cite{R17}. It also accelerates the convergence speed of the tracking error in scenarios where the regressors are constant \cite{R18}. With the growing interest in advancing tuners, we have adapted high-order tuners for graph-related problems \cite{R18,R20}, achieving marginally better outcomes compared to gradient-based methods. Furthermore, we offer insights on designing network weights and selecting constant parameters in the tuners.

This paper makes the following four main contributions. First, we integrate the concept of adaptive control into networked problems with multiple agents, extending its applicability to complex interconnected systems such as star-like, cyclic-like, path, and random networks. Second, we address the challenge of coordinating an arbitrary number of agents with disturbances to follow a designated leader, similar to distributed MRAC. Third, we compare the performance of three distinct tuning algorithms: the gradient descent and two accelerating tuners, providing a comprehensive evaluation of their effectiveness in networked control settings. Finally, 
we not only evaluate the effects of sparsity in subsystem interactions using performance measures (\(L_2-\)norm and \(L_{\infty}-\)norm) across various network configurations and tuners but also demonstrate that our proposed modified high-order tuner significantly outperforms existing tuners, offering novel insights for future research in networked control systems.


We structure the paper as follows. Section \ref{S2} describes the problem formulation of the interconnected systems and the communication network among the agents. In Section \ref{S3}, we outline the networked distributed adaptive control, supported by relevant theorems, and discuss the impact of disturbances. Section \ref{S4} presents three distributed algorithms that utilize both standard gradient descent and modified high-order tuners. In Section \ref{S5}, we deliver the numerical results and findings to explore the effects of network configurations, the number of agents, and the choice of tuners. Finally, Section \ref{S6} highlights our key conclusions and future research.

\textbf{Notations}. $\mathbb{R}^p$ is the $p-$dimensional Euclidean space and $\mathbb{C}^{-}$ refers the open left-half of the complex plane. A symbol $(s)$ shows the complex variable in the Laplace transform. $I_p$ denotes the identity matrix of size $\mathbb{R}^p$ and $P=\diag\{p_i\}$ is the diagonal matrix with entries $p_i,\forall i$. $\mathbf{1}_p=[1,\dots,1]^\top$  is the vector of all ones in $\mathbb{R}^p$. $A\otimes B$ and $A\odot B$ denotes the Kronecker and the Hadamard product between $A$ and $B$. Operator $\tr[A]$, $|A|$, $\lVert A\rVert_2$, and $\lVert A\rVert_F$ define the trace, the absolute value of the element-wise, the Euclidean norm, and the Frobenius norm of matrix $A$.

\section{Preliminaries and Problem Formulation}\label{S2}
\subsection{System Setup}
We consider an interconnected network of subsystems shown in Figure~\ref{F1}, which consists of a leader and $m$ unknown unstable subsystems/agents. Let the unknown subsystems for the followers be defined as follows,
\begin{align} W_i(s)\sim \left\{\begin{aligned}
    \dot{x}_i(t) &= A_ix_i(t) + B_i (u_i(t) + v_i^u), \\
    y_i(t) &= k_{p_i}C_ix_i(t) + v_i^y, 
    \end{aligned}\right. \label{Eq1}
\end{align}
where $x_i \in \mathbb{R}^{n}$ is the state vector, and $u_i, y_i \in \mathbb{R}$ denote the input and output, respectively, for $i=1,2,\dots,m$. The control input $u_i$ and the output measurement $y_i$ are disturbed by the unknown yet constant $v_i^u,v_i^y\in \Omega\subset\mathbb{R}$. The transfer function from $u_i$ to $y_i$ is denoted by $W_i(s)$. The dynamics of the leader as the reference model are written as,
\begin{align} W_\ell(s)\sim \left\{\begin{aligned}
    \dot{x}_\ell(t) &= A_\ell x_\ell(t) + B_\ell r(t), \\
    y_\ell(t) &= k_\ell C_\ell x_\ell(t), 
    \end{aligned}\right. \label{Eq2}
\end{align}
in which $r$ is the reference signal and that is a piecewise-continuous function while $x_\ell\in\mathbb{R}^{n_\ell}$ and $y_\ell\in\mathbb{R}$ represent the reference state and output. Note that $A_i,B_i,C_i,A_\ell,B_\ell,C_\ell$ are constant real matrices with appropriate dimensions whereas $k_{p_i}$ and $k_\ell$ are the high frequency gains. The goal is to design local control input $u_i$ so that the outputs $y_i,\forall i$ follow that of the known stable leader $y_\ell$.
\begin{assumption}\label{assu1}
    The dynamics \eqref{Eq1} are unknown and unstable while \eqref{Eq2} and the signs of $k_{p_i}$ are known. The numerators of $W_i(s),\forall i$ have roots in $\mathbb{C}^-$ while the denominators of $W_i(s),\forall i$ and $W_\ell(s)$ are monic with relative degree $n_d = 1$.
\end{assumption}
\subsection{Communication Network}
We describe the $m$ followers and a leader $\ell$ connected via weighted digraph $\mathcal{G}\coloneqq\{\mathcal{V}=\{1,2,\dots,m \} \cup \{\ell\}, \mathcal{E}, w(\cdot)\}$ where $\mathcal{V}$, $\mathcal{E}$, and $w(\cdot)$ represent the set of agents, directed edges, and the weight function in turn. For simplicity we denote $w(i,j)=w_{ij}$ where $(i,j) \in \mathcal E$. We call the induced subgraph on $m$ followers as $\mathcal{G}_m$, which will also be used interchangeably with the $m$ followers, and the induced subgraph between the leader and the connected followers as $\mathcal{G}_\ell$. Also, we assume that there is a directed path from the leader $\ell$ to all followers.
\begin{figure}[tt]
    \centering
    \includegraphics[width=.55\linewidth]{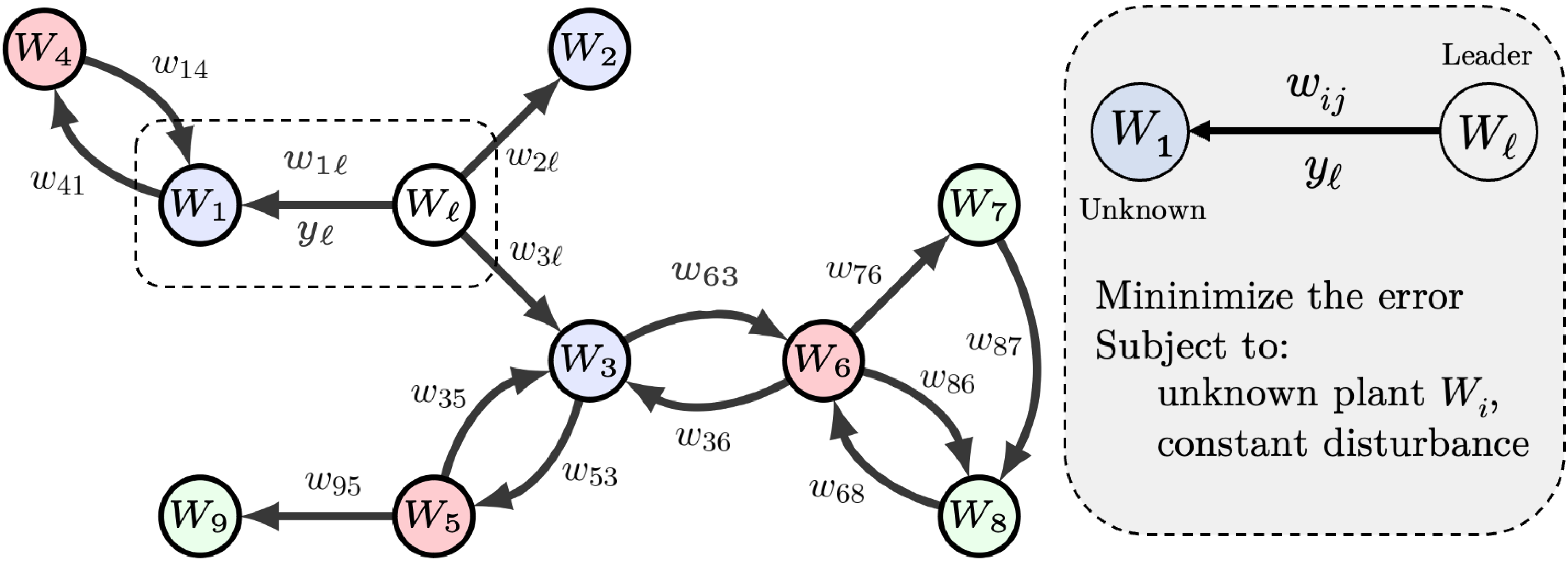}
    \caption{An example of an interconnected network of leader ${\ell}$ and $m= 9$ unknown unstable subsystems/followers.}
    \label{F1}
\end{figure}
The layering colors indicate the $q-$th group of agents from the leader, where the least is the closest, as example shown in Figure~\ref{F1} with $q = \{1,2,3\}$. The incoming arrows for $i-$th system represent the measured neighborhood $j$ with respected weight $w_{ij}$. Note that, the measurement collected to agent $i$ from its neighbors $j\in\mathcal{J}_i$ is designed to be 1, where $w_i=\sum_j w_{ij}=1$, so that the degree matrix for the whole agents is $\mathbb{D}\coloneqq \diag\{D_1,\dots,D_m\} = \diag\{w_1,\dots,w_m\}=I_m$. The measured errors in $\mathcal{G}_m$ are represented as linear operation of its outputs $\Bar{y}$ multiplied by the Laplacian-like matrix of $\mathcal{G}_m$, written as $\mathbb{L}_m\coloneqq \mathbb{D}-\mathbb{A}_m$, and subtracted by the leader $\Bar{y}_\ell$ using $\mathbb{A}_\ell$, with later definition of $\Bar{y}$ and $\Bar{y}_\ell$. The matrices of $\mathbb{L}_m$ and $\mathbb{A}_\ell$ are formulated as follows,
\begin{align*}
    \underbrace{\begin{bmatrix}
        w_1 & -w_{12} & \cdots & -w_{1m} \\
        -w_{21} & w_{2} & \cdots & -w_{2m} \\
        \vdots & \vdots & \ddots & \vdots \\
        -w_{m1} & -w_{m2} & \cdots & w_{m} 
    \end{bmatrix}}_{\mathbb{L}_m\coloneqq\mathbb{D} - \mathbb{A}_m} \textrm{ and } \underbrace{\begin{bmatrix}
        w_{1\ell} & 0 & \cdots & 0 \\
        0 & w_{2\ell} & \cdots & 0 \\
        \vdots & \vdots & \ddots & \vdots \\
        0 & 0 & \cdots & w_{m\ell} 
    \end{bmatrix}}_{\mathbb{A}_\ell|w_{i\ell}\coloneqq 0,\forall q > 1}
\end{align*}
in which, $\mathbb{A}_m$ denotes the adjacency matrix of $\mathcal{G}_m$ whereas $\mathbb{A}_\ell\in\mathbb{R}^{m\times m}=\diag\{w_{1\ell},\dots,w_{m\ell}\}$ is the diagonal matrix containing the weights from the leader to agents in $q=1$. Therefore, the error for agent $i$ is formulated as, 
\begin{align}
    e_i(t)\coloneqq w_iy_i(t) -
    \sum_{j\in\mathcal{J}_i} w_{ij} y_j(t), \label{Eq3}
\end{align}
and the goal is to ensure the boundedness of the errors in $\mathcal{G}_m$, where $\lim_{t\to\infty} \Bar{e}\to 0$, $\Bar{e} = [e_1,\dots,e_m]^\top$, leading to the perfect tracking to the leader $\ell$.
\begin{remark}[Threshold of network]\label{rem1}
    The leader weight matrix is positive semi-definite $\mathbb{A}_\ell\succeq 0$ having the eigenvalues ranging from $0$ to $1$, denoted as $\lambda_i^\ell=[0,1],\forall i=1,\dots,m$, and $\exists\lambda_j^\ell\neq 0$ for $1\leq j\leq m$ in which $\lambda_j^\ell$ is designed depending on the network complexity. The Laplacian-like matrix of $\mathcal{G}_m$ is positive definite $\mathbb{L}_m\succ 0$, ensuring $\lambda_i^m>0,\forall i$.
\end{remark}
\begin{remark}[Communication network]\label{rem2}
    The network is balanced $(\mathbb{L}_m - \mathbb{A}_\ell)\mathbf{1}_{m}=0$. There is always a directed path from the leader $\ell$ to all followers in $\mathcal{G}_m$, otherwise either $\exists\lambda_i^\ell=0,\forall i=1,\dots,m$ or $\exists\lambda_j^m=0$ for $1\leq j\leq m$, violating Remark~\ref{rem1}.
\end{remark}



\section{Distributed Adaptive Control}\label{S3}
We consider the disturbed interconnected systems in \eqref{Eq1} be the following,
\begin{align} \mathbf{W}(s)\sim\left\{\begin{aligned}
    \dot{\Bar{x}}(t) &= \mathbf{A}\Bar{x}(t) + \mathbf{B}\left(\Bar{u}(t) + \nu^u\right), \\
    \Bar{y}(t) &= \mathbf{k}_p\mathbf{C}\Bar{x}(t) + \nu^y,
    \end{aligned}\right.\label{Eq4}
\end{align}
where $\Bar{x} = [x_1^\top,\dots,x_m^\top]^\top\in\mathbb{R}^{\Bar{n}}$ with $\Bar{n}=n\times m$ defines the set of the states, while $\Bar{u} = [u_1,\dots,u_m]^\top\in\mathbb{R}^{m}$ and $\Bar{y} = [y_1,\dots,y_m]^\top\in\mathbb{R}^{m}$ represent the set of inputs and outputs respectively. The matrices 
\begin{align}
    \begin{aligned}
    \mathbf{A} &= \diag\{A_1,\dots,A_m\}\in\mathbb{R}^{\Bar{n}\times\Bar{n}}, & \mathbf{B} &= \diag\{B_1,\dots,B_m\}\in\mathbb{R}^{\Bar{n}\times m}, \\
    \mathbf{C} &= \diag\{C_1,\dots,C_m\}\in\mathbb{R}^{m\times\Bar{n}}, & \mathbf{k}_p &= \diag\{k_{p_1},\dots,k_{p_m}\}\in\mathbb{R}^{m\times m}, 
    \end{aligned} \label{Eq5}
\end{align}
are diagonal blocks of $\mathcal{G}_m$ with high frequency gains $\mathbf{k}_p$. The transfer function version of $\mathcal{G}_m$ is denoted as $\mathbf{W}(s) = \diag\{W_1(s),\dots,W_m(s)\}$. We design so that the persistent excitation of $\Bar{\nu}=[\nu^{u\top},\nu^{y\top}]^\top$, $\nu^\alpha = [v_1^\alpha,\dots,v_m^\alpha]^\top$, with $\alpha=\{u,y\}$ are less than that of the reference $\Bar{r}$, where $\sup(\Bar{\nu}) < \Bar{r}$. We also expand the leader $\ell$ in \eqref{Eq2} as follows,
\begin{align} \mathbf{W}_\ell(s)\sim\left\{\begin{aligned}
    \dot{\Bar{x}}_\ell(t) &= \mathbf{A}_\ell\Bar{x}_\ell(t) + \mathbf{B}_\ell\Bar{r}(t), \\
    \Bar{y}_\ell(t) &= \mathbf{k}_\ell\mathbf{C}_\ell\Bar{x}_\ell(t),
    \end{aligned}\right.\label{Eq6}
\end{align}
where $\Bar{x}_\ell = \mathbf{1}_{m}\otimes x_\ell$, $\Bar{y}_\ell  = \mathbf{1}_m \otimes y_\ell$ and $\Bar{r} = \mathbf{1}_m \otimes r$ denote the set of states, outputs and references of the leader $\ell$ while $\mathbf{A}_\ell$, $\mathbf{B}_\ell$, and $\mathbf{C}_\ell$ are the diagonal matrices in the forms of, 
\begin{align}
    \mathbf{A}_\ell\coloneqq I_m\otimes A_\ell, \qquad \mathbf{B}_\ell\coloneqq I_m\otimes B_\ell, \qquad \mathbf{C}_\ell\coloneqq I_m\otimes C_\ell, \label{E7}
\end{align}
with the similar dimensions to $\mathbf{A}$, $\mathbf{B}$, and $\mathbf{C}$ respectively. Likewise, the transfer function of the leader $\ell$ is defined as $\mathbf{W}_\ell(s) = I_m\otimes W_\ell(s)$ with high frequency gains $\mathbf{k}_\ell = I_m\otimes k_\ell$. Note that, the interactions among agents $\mathcal{G}_m$ and the leader $\ell$ are constructed based on $\mathbb{L}_m$ and $\mathbb{A}_\ell$. 

It is obvious that if $\mathcal{G}_m$ is known, then the control input $\Bar{u}$ satisfying $\lim_{t\to\infty} \Bar{e}\coloneqq[e_1,\dots,e_m]^\top = 0$ is to choose $\Bar{u}=\mathbf{M}(s)(\mathbf{1}_m \otimes r)$, where $\mathbf{M}(s)=\mathbf{W}_\ell(s)\mathbf{W}^{-1}(s)$. However, for the unknown $\mathcal{G}_m$, the engineering for unknown constant $\mathbf{k}_p$, the zeros and the poles of $\mathbf{W}(s)$ is required to be solved. Here, we construct the problem into two,
\begin{enumerate}
    \item the unknown $\mathbf{k}_p$ of $\mathbf{W}(s)$
    \item the unknown $\mathbf{k}_p$, zeros and poles of $\mathbf{W}(s)$
\end{enumerate}
Regarding the first part, we assume $\mathbf{W}(s)=\mathbf{k}_p\mathbf{W}_\alpha(s)$ and $\mathbf{W}_\ell(s)=\mathbf{k}_\ell\mathbf{W}_\alpha(s)$, where $\mathbf{W}_\alpha(s)$ is the transfer function of $\mathbf{A}_\ell,\mathbf{B}_\ell,\mathbf{C}_\ell$. The optimal estimate for the unknown $\mathbf{k}_p$ is $\mathbf{k}^\ast = \mathbf{k}_\ell\mathbf{k}_p^{-1}$ and with control input $\Bar{u} = (\mathbf{k}^\ast + \Tilde{\mathbf{k}})(\mathbf{1}_m \otimes r)$, then the tracking errors $\Bar{e}$ for the unknown $\mathbf{k}_p$ only of $\mathcal{G}_m$ are formulated as follows,
\begin{align}
    \Bar{e}(t) &= \mathbb{L}_m\Bar{y}(t) - \mathbb{A}_\ell\Bar{y}_\ell(t)\nonumber\\
    &= \left[\mathbb{L}_m\mathbf{k}_p\mathbf{W}_\alpha(s)(\mathbf{k}^\ast + \Tilde{\mathbf{k}}) - \mathbb{A}_\ell\mathbf{k}_\ell\mathbf{W}_\alpha(s)\right](\mathbf{1}_m \otimes r(t)) \nonumber\\
    &= \mathbb{L}_m\mathbf{k}_p\mathbf{W}_\alpha(s)\Tilde{\mathbf{k}}(\mathbf{1}_m \otimes r(t)), \label{Eq8}
\end{align}
and due to $\mathbb{D}=I_m\eqqcolon w_i,\forall i$, then $(\mathbb{L}_m-\mathbb{A}_\ell)\mathbf{1}_m=0$. To deal with the second case, we need the 
Meyer-Kalman-Yakubovic (MKY) lemma as a basis to generate the adaptive laws and guarantee the stability.  
\begin{lemma}\label{lem1}
    Consider the networked system in \eqref{Eq4} where the pairs of $(\mathbf{A},\mathbf{B})$ and $(\mathbf{A},\mathbf{C})$ are stabilizable and detectable, assuming the strictly positive realness of the transfer function $\mathbf{W}_\beta(s)\triangleq \mathbf{C}(sI_{\Bar{n}}-\mathbf{A})^{-1}\mathbf{B}$. Moreover, let the controller be $\Bar{u}\coloneqq\Theta^\top(t)\Bar{\eta}(t)$ where $\eta_i:\mathbb{R}^+\to\mathbb{R}^{p}$, $\Bar{\eta} = [\eta_1^\top,\dots,\eta_m^\top]^\top$, and $\Bar{y}:\mathbb{R}^+\to\mathbb{R}^m$ be the measured time-varying functions while $\Theta\in\mathbb{R}^{\Bar{p}\times m}$ with $\Bar{p}=p\times m$ be the adaptive term of the form,
    \begin{align}
        \dot{\Theta}^\top(t) = -\sign(\mathbf{k}_p)\Bar{y}(t)\Bar{\eta}^\top(t),
         \label{Eq9}
    \end{align}
    then the equilibrium $(\Bar{x},\Theta)\coloneqq 0$ is uniformly stable in the large.
\end{lemma}
\begin{proof}
    Since $\mathbf{W}_\beta(s)$ is strictly positive real (SPR), then $\exists Q=Q^\top\succ 0, P=P^\top\succ 0$ such that,
    \begin{align}
        \mathbf{A}^\top P + P\mathbf{A} = -Q, \quad P\mathbf{B} = \mathbf{C}^\top, \label{Eq10}
    \end{align}
    and choosing the positive definite Lyapunov function of $V(\Bar{x},\Theta)>0$ leads to the negative semi-definite function of its time derivative $\dot{V}(\Bar{x},\Theta)\leq 0$, as written in the following,
    \begin{align*}
        V(\Bar{x},\Theta) &= \Bar{x}^\top(t)P\Bar{x}(t) + \tr\left[\Theta(t)|\mathbf{k}_p^{-1}|\Theta^\top(t)\right] \\
        \dot{V}(\Bar{x},\Theta) &= \Bar{x}^\top(t)\left[\mathbf{A}^\top P + P\mathbf{A}\right]\Bar{x}(t) + 2\Bar{x}^\top(t) P\mathbf{B}\Theta^\top(t)\Bar{\eta}(t) - 2\tr\left[\Bar{\eta}(t)\Bar{y}^\top(t)|\mathbf{k}_p^{-1}|\Theta^\top(t)\right] \\
        &= -\Bar{x}^\top(t)Q\Bar{x}(t) \leq 0, 
    \end{align*}
    considering \eqref{Eq10} and if we choose $\dot{\Theta}^\top$ as in \eqref{Eq9}. Note that, it is required to show that $\mathbf{W}_\beta(s)$ be SPR.
\end{proof}
Now, for the unknown $\mathbf{k}_p$, zeros and poles of $\mathcal{G}_m$, let us define $(\mathbf{N}(s),\mathbf{D}(s)),(\mathbf{N}_\ell(s),\mathbf{D}_\ell(s))$ be the diagonal matrices containing the set of numerators and denominators of $\mathcal{G}_m$ and the leader $ \ell$ in turn, 
\begin{align*}
    \mathbf{N}(s) &= \diag\{n_1(s),\dots,n_m(s)\}, &\mathbf{N}_\ell(s) &= I_m\otimes n_\ell(s), \\
    \mathbf{D}(s) &= \diag\{d_1(s),\dots,d_m(s)\}, & \mathbf{D}_\ell(s) &= I_m\otimes d_\ell(s), 
\end{align*}
where the transfer functions are $\mathbf{W}(s) = \mathbf{k}_p\mathbf{N}(s)\mathbf{D}^{-1}(s)$ and $\mathbf{W}_\ell(s) = \mathbf{k}_\ell\mathbf{N}_\ell(s)\mathbf{D}_\ell^{-1}(s)$. The feed-forward and feedback mechanism to adjust the unknown $\mathbf{N}(s)$ and the unknown $\mathbf{D}(s)$ are written as follows, 
\begin{align}
    \Psi_d^{\ast\top}\mathbf{H}\left[\Bar{u}(t)+\nu^u\right] &= \Psi_d^{\ast\top} \Bar{z}(t), \label{Eq11}\\
    \left[\Phi_d^{\ast\top}\mathbf{H} + T_d^{\ast}\right]\Bar{y}(t) &= \Phi_d^{\ast\top}\Bar{\omega}(t) + T_d^{\ast}\Bar{y}(t), \label{Eq12}
\end{align}
where the optimal gain matrices of $\Psi_d^{\ast\top}=\diag\{\psi^{\ast\top}_1,\dots,\psi^{\ast\top}_m\}$, $\Phi_d^{\ast\top}=\diag\{\phi^{\ast\top}_1,\dots,\phi^{\ast\top}_m\}$, and $T_d^\ast=\diag\{\tau_1^\ast,\dots,\tau_m^\ast\}$ show the adaptive terms with $\psi_i^\ast,\phi_i^\ast\in\mathbb{R}^{n-1}$, $\tau_i^\ast\in\mathbb{R}, \forall i$. The known systems $\mathbf{H}(s)$ are defined as,
\begin{align}
    \mathbf{H}(s) = (sI_{m(n-1)} - \Lambda_{m(n-1)})^{-1} \vartheta_{m(n-1)} ,\label{Eq13}
\end{align}
where the matrices of $\Lambda_{m(n-1)}=I_m\otimes\Lambda_{n-1}$ and $\vartheta_{m(n-1)}=I_m\otimes\vartheta_{n-1}$ are the stable systems where the pair $(\Lambda_{n-1},\vartheta_{n-1})$ is of order $n-1,\forall i$. Note that, $\Bar{z}\in\mathbb{R}^{\Bar{q}}=[z_1^\top,\dots,z_m^\top]^\top$ and $\Bar{\omega}\in\mathbb{R}^{\Bar{q}}=[\omega_1^\top,\dots,\omega_m^\top]^\top$ with $\Bar{q}=(n-1)\times m$ can be also represented as
\begin{align}
    \dot{\Bar{z}}(t) &= \Lambda_{m(n-1)}\Bar{z}(t) + \vartheta_{m(n-1)}\left[\Bar{u}(t)+\nu^u\right], \label{Eq14}\\ 
    \dot{\Bar{\omega}}(t) &= \Lambda_{m(n-1)}\Bar{\omega}(t) + \vartheta_{m(n-1)}\Bar{y}(t),
    \label{Eq15}
\end{align}
therefore, the control $\Bar{u}$ is then defined as,
\begin{align}
    \Bar{u}(t) &= \Theta^\top(t)\Bar{\eta}(t), \label{Eq16a}
\end{align}
where $\Bar{\eta} = [\eta_1^\top,\dots,\eta_m^\top]^\top$, $\eta_i =[r_i,z_i^\top,\omega_i^\top,y_i]^\top, \forall i = 1,\dots,m$. Now, we need to show that $\Bar{e}, \Bar{z}, \Bar{w}$ are bounded such that by considering \eqref{Eq4}, \eqref{Eq11}-\eqref{Eq16a} and the parameter error $\Theta = \Theta^\ast + \Tilde{\Theta}$, the outputs of $\mathcal{G}_m$ are denoted as,
\begin{align}
    \dot{\Bar{x}}_a(t) = \mathbf{A}_a\Bar{x}_a(t) + \mathbf{B}_a\left[\Tilde{\Theta}^{\top}(t)\Bar{\eta}(t) + \mathbf{k}^\ast(t)\Bar{r}(t)\right], \quad \Bar{y}(t) = \mathbf{C}_a\Bar{x}_a(t), \label{Eq17}
\end{align}
where $\Bar{x}_a = [\Bar{x}^{\top} ~ \Bar{z}^{\top} ~ \Bar{\omega}^{\top}]^\top$ and,
\begin{align}\begin{aligned}
    \mathbf{A}_a &= \begin{bmatrix}
        \mathbf{A}+\mathbf{B}T_d^{\ast}\mathbf{k}^\ast\mathbf{C} & \mathbf{B}\Psi_d^{\ast\top} & \mathbf{B}\Phi_d^{\ast\top} \\
        \vartheta T_d^{\ast}\mathbf{k}^\ast\mathbf{C} & \Lambda + \vartheta\Psi_d^{\ast\top} & \vartheta\Phi_d^{\ast\top} \\
        \vartheta\mathbf{k}^\ast\mathbf{C} & 0 & \Lambda
    \end{bmatrix}, \\ 
    \mathbf{B}_a &= \begin{bmatrix}
        \mathbf{B}^\top & \vartheta^\top & 0
    \end{bmatrix}^\top, \quad \mathbf{C}_a = \begin{bmatrix}
        \mathbf{k}^\ast\mathbf{C} & 0 & 0
    \end{bmatrix}. \end{aligned} \label{Eq18}
\end{align}
It also follows that the leader $ \ell$ can be constructed using the optimal gains of $\mathbf{k}^\ast$, $\Psi_d^{\ast}$, $\Phi_d^{\ast}$, and $T_d^{\ast}$, such that \eqref{Eq19} is equal to $\Bar{y}_\ell\coloneqq \mathbf{k}_\ell\mathbf{C}_\ell(sI_{\Bar{n}} - \mathbf{A}_\ell)^{-1}\mathbf{B}_\ell\Bar{r}$, therefore
\begin{align}
    \dot{\Bar{x}}_a^\ast(t) = \mathbf{A}_a\Bar{x}_a^\ast(t) + \mathbf{B}_a \mathbf{k}^\ast(t)\Bar{r}(t), \quad \Bar{y}_\ell(t) = \mathbf{C}_a\Bar{x}_a^\ast(t), \label{Eq19}
\end{align}
in which by considering the state error $\Bar{e}_a = \hat{\mathbb{L}}_m\Bar{x}_a - \hat{\mathbb{A}}_\ell\Bar{x}_a^\ast$ and the output error $\Bar{e} \coloneqq \mathbb{L}_m\Bar{y} - \mathbb{A}_\ell(\mathbf{1}_m\otimes y_\ell)$, where $\hat{\mathbb{L}}_m=\diag\{(\mathbb{L}_m\otimes I_n),I_{\Bar{q}},I_{\Bar{q}}\}$ and $\hat{\mathbb{A}}_\ell=\diag\{(\mathbb{A}_\ell\otimes I_n),I_{\Bar{q}},I_{\Bar{q}}\}$, then 
\begin{subequations}
\begin{align}
    \dot{\Bar{e}}_a(t) &= \hat{\mathbb{L}}_m\mathbf{A}_a\Bar{x}_a(t) + \hat{\mathbb{L}}_m\mathbf{B}_a\left[\Tilde{\Theta}^{\top}(t)\Bar{\eta}(t) + \mathbf{k}^\ast(t)\Bar{r}(t)\right] - \hat{\mathbb{A}}_\ell\mathbf{A}_a\Bar{x}_a^\ast(t) - \hat{\mathbb{A}}_\ell\mathbf{B}_a \mathbf{k}^\ast(t)\Bar{r}(t), \nonumber\\
    &= \mathbf{A}_a\Bar{e}_a(t) + \hat{\mathbb{L}}_m\mathbf{B}_a\Tilde{\Theta}^{\top}(t)\Bar{\eta}(t) \label{Eq20a}\\
    \Bar{e}(t) &= \mathbb{L}_m\mathbf{C}_a\Bar{x}_a(t) -  \mathbb{A}_\ell\mathbf{C}_a\Bar{x}_a^\ast(t) \label{Eq20b}\\
    &= \mathbb{L}_m\mathbf{C}_a(sI_{a} - \mathbf{A}_a)^{-1}\mathbf{B}_a\left[\Tilde{\Theta}^{\top}(t)\Bar{\eta}(t) + \mathbf{k}^\ast(t)\Bar{r}(t)\right] -\mathbb{A}_\ell\mathbf{C}_a(sI_{a} - \mathbf{A}_a)^{-1}\mathbf{B}_a\mathbf{k}^\ast(t)\Bar{r}(t) \nonumber\\
    &= \mathbb{L}_m\mathbf{C}_a(sI_{a} - \mathbf{A}_a)^{-1}\mathbf{B}_a\Tilde{\Theta}^{\top}(t)\Bar{\eta}(t) \eqqcolon \mathbf{C}_a\Bar{e}_a(t). \label{Eq20c}
\end{align}
\end{subequations}
We can generate the adaptive laws using Lemma~\ref{lem1} in which $\exists Q_a=Q_a^\top\succ 0, P_a=P_a^\top\succ 0$ such that, $\mathbf{A}_a^\top P_a + P_a\mathbf{A}_a = -Q_a$, and $P_a\mathbf{B}_a = \mathbf{C}_a^\top$ since $\mathbf{W}_\gamma(s)\triangleq \mathbf{C}_a(sI_{a}-\mathbf{A}_a)^{-1}\mathbf{B}_a$ is SPR given $\mathbf{W}_\gamma(s)=\mathbf{k}_p\mathbf{k}_\ell^{-1}\mathbf{W}_\ell(s)$. The stability can be guaranteed from the following Lyapunov function,
\begin{subequations}
\begin{align}
    V(\Bar{e}_a,\Tilde{\Theta}) &= \Bar{e}_a^\top(t)P\Bar{e}_a(t) + \tr\left[\Tilde{\Theta}(t)\Gamma_a^{-1}\Tilde{\Theta}^\top(t)\right] \label{Eq21a}\\
    \dot{V}(\Bar{e}_a,\Tilde{\Theta}) &= \Bar{e}_a^\top(t)\left[\mathbf{A}_a^\top P_a + P_a\mathbf{A}_a\right]\Bar{e}_a(t) + 2\Bar{e}_a^\top(t) P_a\hat{\mathbb{L}}_m \mathbf{B}_a\Tilde{\Theta}^\top(t)\Bar{\eta}(t) \nonumber\\
    &\quad- 2\tr\left[\Bar{\eta}(t)\Bar{e}^\top(t)\mathbb{L}_m\Gamma_a\Gamma_a^{-1}\Tilde{\Theta}^\top(t)\right] \label{Eq21b}\\
    &= -\Bar{e}_a^\top(t)Q_a\Bar{e}_a(t) \leq 0, \label{Eq21c}
\end{align}
\end{subequations}
if we design the adaptive laws as follows,
\begin{align}
    \dot{\Tilde{\Theta}}^\top(t) = \dot{\Theta}^\top(t) = -\sign(\mathbf{k}_p)\Gamma_a\mathbb{L}_m^\top\Bar{e}(t)\Bar{\eta}^\top(t).
    \label{Eq22}
\end{align}
\begin{remark}\label{rem3}
    The matrix $\mathbf{C}_a$ makes $\mathbb{L}_m$ in \eqref{Eq20c} sufficient to capture $\Bar{e}_a(t)$ containing $\hat{\mathbb{L}}_m$ in \eqref{Eq20a}. Also, the cancellation in \eqref{Eq21b} occurs due to the fact $\Bar{e}_a^\top(t)P_a\hat{\mathbb{L}}_m \mathbf{B}_a = \Bar{e}_a^\top(t) \hat{\mathbb{L}}_m \mathbf{C}_a^\top = \Bar{e}^\top(t)\mathbb{L}_m$, considering $P_a\mathbf{B}_a = \mathbf{C}_a^\top$. Given $n_d=1$ in Assumption~\ref{assu1}, since $\mathbf{H}(s)$ is stable and strictly proper, then $\Bar{\nu}$ is cancelled out and \eqref{Eq17} is valid. It is due to a pole in the feedforward transfer function at $s=0$ for each subsystem where the disturbance-term decays exponentially to zero.
\end{remark}
Furthermore, we provide one lemma and two theorems such that the outputs \eqref{Eq4} and the errors $\Bar{e}$ of the balanced connected networks are bounded while \eqref{Eq17} has a solution and guarantees the tracking of the leader $ \ell$. 

\begin{lemma}\label{lem2}
    Let the stable systems in \eqref{Eq13} be shown as $\mathbf{H}(s)\coloneqq\mathbf{N}_\Lambda(s)\mathbf{D}_{\Lambda}^{-1}(s)$, where $\mathbf{N}^\top_\Lambda(s) = \diag\{n_{\lambda_1}^\top(s),\dots,n_{\lambda_{m}}^\top(s)\}$ and $\mathbf{D}_\Lambda(s) = \diag\{d_{\lambda_1}(s),\dots,d_{\lambda_{m}}(s)\}$. There exist the optimal gains $\mathbf{k}^\ast$, $\Psi_d^{\ast}$, $\Phi_d^{\ast}$, and $T_d^{\ast}$, so that the following matching condition is achieved, where
    \begin{align}\begin{aligned}
        \Psi_d^{\ast\top}\mathbf{N}_\Lambda\mathbf{D} + \left(\Phi_d^{\ast\top}\mathbf{N}_\Lambda + T_d^\ast\mathbf{D}_\Lambda\right)\mathbf{k}_p\mathbf{N} &= \mathbf{D}_\Lambda\left[\mathbf{D} - \mathbf{k}^\ast\left(\mathbf{k}_\ell\mathbf{N}_\ell\mathbf{D}_\ell^{-1}\right)^{-1}\mathbf{k}_p\mathbf{N}\right].
        \end{aligned} \label{Eq23}
    \end{align}
\end{lemma}
\begin{proof}\label{pro2}
    Let $n=n_\ell$, $\mathbf{k}^\ast=\mathbf{k}_p^{-1}$ and $\Pi(s)\coloneqq \mathbf{D}_\Lambda(\mathbf{D} - \mathbf{W}_\ell^{-1}\mathbf{N})$, then $\Pi(s) = \diag\{\pi_1(s),\dots,\pi_m(s)\}$, in which each entry of $\Pi(s)$ is denoted as $\pi_i(s) = d_{\lambda_i}(s)\left[d_i(s) - d_{\ell_i}(s)n_i(s)/k_\ell n_{\ell_i}(s)\right]$ showing that $\pi_i(s)$ is non-monic of order $2n-2$.
    Let us define $\mathbf{N}_\Lambda$ having all ones coefficients, then 
    \begin{align*}
        \Psi_d^{\ast\top}\mathbf{N}_\Lambda(s)=\Psi^{\ast}(s), \quad \Phi_d^{\ast\top}\mathbf{N}_\Lambda(s)=\Phi^{\ast}(s), 
    \end{align*}
    where $\Psi^{\ast}(s)=\diag\{\psi_1(s),\dots,\psi_m(s)\}$ and also $\Phi^{\ast}(s)=\diag\{\phi_1(s),\dots,\phi_m(s)\}$. Now supposed,
    \begin{align*}
        T_d^{\ast}\mathbf{D}_\Lambda = \diag\{\tau_1d_{\lambda_1}(s),\dots,\tau_m d_{\lambda_m}(s)\} \eqqcolon T^{\ast}(s),
    \end{align*}
    and by combining $\mathbf{D}$ and $\mathbf{k}_p\mathbf{N}$ in series, we have
    \begin{align}
        \mathbb{S}(s)\Theta_s^\ast(s)\coloneqq\begin{bmatrix}
            \mathbf{D} & \mathbf{k}_p\mathbf{N}
        \end{bmatrix}\begin{bmatrix}
            \Psi^\ast(s) \\
            \Phi^\ast(s) + T^{\ast}(s)
        \end{bmatrix} = \Pi(s). \label{Eq24}
    \end{align}
    If we expand each Laplace variables of $\mathbb{S}(s)$, $\Theta_s^\ast(s)$ and $\Pi(s)$ into vector element, e.g. one of the elements in $\Pi(s)$, $\pi_1(s) \coloneqq c_ns^n + c_{n-1}s^{n-1}+\dots+c_0\rightarrow \Bar{\pi}_1 = [c_n, c_{n-1},\dots,c_0]^\top$, defined as $\Bar{\mathbb{S}}$, $\Bar{\Theta}_s^\ast$, and $\Bar{\Pi}$ in turn, then 
    \begin{align}
        \Bar{\mathbb{S}}\Bar{\Theta}_s^\ast = \Bar{\Pi} \longrightarrow \begin{cases}
            \Bar{\mathbb{S}} &= \diag\{\Bar{s}_1,\dots,\Bar{s}_m\} \in\mathbb{R}^{r_m\times r_m}\\
            \Bar{\Theta}_s^\ast &= \diag\{\Bar{\xi}_1,\dots,\Bar{\xi}_m\} \in\mathbb{R}^{r_m}\\
            \Bar{\Pi} &= \diag\{\Bar{\pi}_1,\dots,\Bar{\pi}_m\} \in\mathbb{R}^{r_m}
        \end{cases}, \label{Eq25}
    \end{align} 
    where $\Bar{\mathbb{S}}\in\mathbb{R}^{r_m\times r_m}$ with $r_m=(2n-1)m$, while $\Bar{s}_i$ and $\Bar{\xi}_i$ are formulated as, 
    \begin{align*} 
        \Bar{s}_i &= \begin{bmatrix}
            \Bar{s}_i^1 | k_{p_i}\Bar{s}_i^2
        \end{bmatrix} = \begin{cases}
            \Bar{s}_i^1 = \begin{bmatrix}
            [d_i,0\otimes\mathbf{1}^\top_k]^\top & [0,d_i,0\otimes\mathbf{1}^\top_{k-1}]^\top & \dots & [0\otimes\mathbf{1}^\top_k,d_i]^\top
        \end{bmatrix}\in\mathbb{R}^{2n-1\times n-1} \\ \noalign{\vskip 5pt}
        \Bar{s}_i^2 = \begin{bmatrix}
            [n_i,0\otimes\mathbf{1}^\top_{k+1}]^\top & [0,n_i,0\otimes\mathbf{1}^\top_{k}]^\top & \dots &[0\otimes\mathbf{1}^\top_{k+1},n_i]^\top
        \end{bmatrix}\in\mathbb{R}^{2n-1\times n} 
        \end{cases}\\
        \Bar{\xi}_i &= [\psi_i^\top, ([0, \phi_i^\top]^\top + \tau_id_{\lambda_i})^\top]^\top \in\mathbb{R}^{2n-1},
    \end{align*}
   where $k=n-2$. If $\mathbf{N}(s)$ and $\mathbf{D}(s)$ are coprime, then the rank of $\mathbb{S}$, defined as $\rho(\mathbb{S})$, be full rank and \eqref{Eq25} has a solution, otherwise the solution is not unique. However, given $\mathbf{N}(s)=\Bar{\mathbf{N}}(s)\mathbf{P}(s)$ and $\mathbf{D}(s)=\Bar{\mathbf{D}}(s)\mathbf{P}(s)$, for some polynomial $\mathbf{P}(s)$, such that $\Bar{\mathbf{N}}(s)$ and $\Bar{\mathbf{D}}(s)$ are coprime, and with the help of $\mathbf{Q}_1,\mathbf{Q}_2,\mathbf{Q}_3$ and some formulation as in Appendix~\ref{A2}, then there exist $\Theta^\ast$ such that \eqref{Eq23} is satisfied.
\end{proof}
\begin{theorem}\label{thm1}
    Given $\mathbf{k}^\ast$, $\Psi_d^{\ast}$, $\Phi_d^{\ast}$, and $T_d^{\ast}$ satisfying \eqref{Eq23} and $\Theta\equiv\Theta^\ast$ in \eqref{Eq22}, then the controller $\Bar{u}=\Theta^{\ast\top}\Bar{\eta}$ ensures the boundedness of all the signals in the closed-loop form and
    \begin{align}
        \Bar{e} \coloneqq \mathbb{L}_m\Bar{y}(t) - \mathbb{A}_\ell\Bar{y}_\ell(t) = \mathbf{1}_m\otimes\epsilon_0, \label{Eq26}
    \end{align}
    where $\epsilon_0$ denotes the exponentially decaying initial condition.
\end{theorem}
\begin{proof}
    Considering the stable systems of both $\mathbf{D}_\Lambda$ and $\mathbf{N}$, and operating both sides of $\eqref{Eq23}$ with $\mathbb{L}_m\Bar{y}$, after some algebra shown in Appendix~\ref{A3} then
    \begin{align}
        \mathbb{L}_m\Bar{u}(t) &= \mathbb{L}_m\Psi_d^{\ast\top}\mathbf{H}\Bar{u}(t) + \mathbb{L}_m\Phi_d^{\ast\top}\mathbf{H}\Bar{y}(t) + \mathbb{L}_m T_d^\ast\Bar{y}(t) + \mathbb{L}_m\mathbf{k}^\ast\mathbf{W}_\ell^{-1}\Bar{y}(t) + (\mathbf{1}_m\otimes\epsilon_1),  \label{Eq27}
    \end{align}
    and let $\Theta\equiv\Theta^\ast$, then $\Bar{u}=\Theta^{\ast\top}\Bar{\eta}$, such that by adding $(\mathbb{L}_m-\mathbb{A}_\ell)\mathbf{k}^\ast\Bar{r}\coloneqq 0$ into \eqref{Eq27}, we have
    \begin{align}
        \mathbf{k}^\ast\mathbf{W}_\ell^{-1}(s)\left[\mathbb{L}_m\Bar{y}(t) - \mathbb{A}_\ell\Bar{y}_\ell(t)\right] + (\mathbf{1}_m\otimes\epsilon_1) = 0.  \label{Eq28}
    \end{align}
    Since $\mathbf{W}_\ell$ is stable and using \eqref{Eq4}, then $\Bar{y}\in\mathcal{L}^\infty$, $\Bar{u}\in\mathcal{L}^\infty$, while $\Bar{z}$ and $\Bar{\omega}$ are bounded.
\end{proof}
Let $\Gamma_a\in\mathbb{R}^{m\times m}\succ 0$ and $Q_a\succ 0$ be symmetric with $P_a=P_a^\top\succ 0$ be the solution of the Lyapunov functions $\mathbf{A}_a^\top P_a + P_a\mathbf{A}_a = -Q_a$ and $P_a\mathbf{B}_a = \mathbf{C}_a^\top$ then the following theorem, elaborated from Lemma~\ref{lem1}, holds. 
\begin{theorem}\label{thm2}
    Consider the networked system \eqref{Eq4} of $\mathcal{G}_m$ and \eqref{Eq6} of the leader $\ell$ with the Laplacian-like $\mathbb{L}_m$ and the leader weight $\mathbb{A}_\ell$ satisfying Remark~\ref{rem1} and \ref{rem2} along with the disturbance-term in Remark~\ref{rem3}. The pair $(\mathbf{A},\mathbf{B})$ is stabilizable satisfying Assumption~\ref{assu1} and let the controller be $\Bar{u}\coloneqq\Theta^\top(t)\Bar{\eta}(t)$ where $\eta_i:\mathbb{R}^+\to\mathbb{R}^{p}$, $\Bar{\eta} = [\eta_1^\top,\dots,\eta_m^\top]^\top$, and $\Bar{y}:\mathbb{R}^+\to\mathbb{R}^m$ be the measured time-varying functions while $\Theta\in\mathbb{R}^{\Bar{p}\times m}$ with $\Bar{p}=p\times m$ be the adaptive term of the form \eqref{Eq22}, then the boundedness of $\Bar{e},\Bar{z},\Bar{\omega}$ in $\mathcal{G}_m$ is guaranteed, leading to the asymptotic tracking to the leader $\ell$.
\end{theorem}
\begin{proof}
    Using \eqref{Eq4}, \eqref{Eq14}, \eqref{Eq15}, \eqref{Eq16a}, and the parameter error $\Theta = \Theta^\ast + \Tilde{\Theta}$, then \eqref{Eq17} is obtained to describe $\mathcal{G}_m$ while the leader $\ell$ is defined in \eqref{Eq19}, showing the perfect matching of \eqref{Eq4}-\eqref{Eq6}, using $\Bar{u}^\ast$ as in \eqref{Eq23}. Considering the errors of $\Bar{e}_a = \hat{\mathbb{L}}_m\Bar{x}_a - \hat{\mathbb{A}}_\ell\Bar{x}_a^\ast$ and $\Bar{e} \coloneqq \mathbb{L}_m\Bar{y} - \mathbb{A}_\ell\Bar{y}_\ell$, then \eqref{Eq20a}-\eqref{Eq20c} is bounded, proven by \eqref{Eq21a}-\eqref{Eq21c} if the adaptive law in \eqref{Eq22} is chosen.
\end{proof}

\section{Distributed High-Order Tuners}\label{S4}
We discuss two common errors in adaptive system, the tracking error $\Bar{e}$ between the leader $\ell$ and the interconnected system $\mathcal{G}_m$ and the parameter estimation error $\Tilde{\Theta} = \Theta - \Theta^\ast$. We propose two high-order tuners, $\Theta_1,\Theta_2\in\mathbb{R}^{mp_i\times m}$ inspired by \cite{R18}, against the gradient-based tuner in \eqref{Eq22}. The two tuners come from the Bregman Lagrangian in the form of,
\begin{align}
    \mathscr{L}(\Theta_j^\top,\dot{\Theta}_j^\top,t) = e^{\Bar{\alpha}_j-\Bar{\gamma}_j}\left[\mathbb{D}_b(f(\Theta_j^\top(t)),\Theta_2^\top(t))-e^{\Bar{\beta}_j} L(\Theta_j^\top(t))\right], \label{Eq29}
\end{align}
where $f(\Theta_j^\top)=\Theta_j^\top+e^{-\Bar{\alpha}_j}\dot{\Theta}_j^\top$ and the Bregman divergence is denoted as $\mathbb{D}_b(y,x)=b(y)-b(x)-\tr\left[(y-x)\nabla b(x)^\top\right]$ with $b(x)=0.5\lVert x\rVert^2_F$ for all $j=1,2$. Moreover, $L(\cdot)$ defines the time-varying loss function from [20a], where
\begin{align}
    L(\cdot) = \frac{1}{2}\left(\frac{d}{dt}\Bar{e}_a^\top(t) P_a\Bar{e}_a(t) + \Bar{e}_a^\top(t) Q_a\Bar{e}_a(t)\right), \label{Eq30}
\end{align}
resulting the update laws for specific $\dot{\Theta}_j^\top$ as $\Gamma_\gamma\nabla_{\Theta_j}L(\Theta_j)$.
Given $\Gamma_\gamma\coloneqq\gamma I_m,\Gamma_\beta\coloneqq\beta I_m\in\mathbb{R}^{m\times m}\succ 0$ for simplicity such that $\tr\left[\Gamma_\gamma\right]\coloneqq\gamma\times m=\gamma_m$, $\tr\left[\Gamma_\beta\right]\coloneqq\beta\times m=\beta_m$ and the normalization $\mathcal{N} = 1 + \mu\Bar{\eta}^\top\Bar{\eta}$, such that by substituting,
\begin{subequations}
\begin{align}
    \Bar{\alpha}_1&=\ln(\Gamma_\beta\mathcal{N}^{-1}),& \Bar{\beta}_1&=\ln(\Gamma_\gamma\Gamma_\beta^{-1}\mathcal{N}^{-1}),&  \Bar{\gamma}_1&=\int_{t_0}^t\Gamma_\beta\mathcal{N}\;ds, \label{Eq31a}\\
    \Bar{\alpha}_2&=0,& \Bar{\beta}_2&=\ln(\Gamma_\gamma\Gamma_\beta\mathcal{N}^{-1}),& \Bar{\gamma}_2&=\Gamma_\beta(t-t_0), \label{Eq31b}
\end{align}
\end{subequations}
then we have,
\begin{subequations}
\begin{align}
    \mathscr{L}(\Theta_1^\top,\dot{\Theta}_1^\top,t) &= e^{\Bar{\gamma}_1}\left(\frac{1}{2}\Gamma_\beta^{-1}\mathcal{N}^{-1}\lVert\dot{\Theta}_1(t)\rVert^2_F - \Gamma_\gamma L(\Theta_1(t))\right), \label{Eq32a}\\
    \mathscr{L}(\Theta_2^\top,\dot{\Theta}_2^\top,t) &= e^{\Bar{\gamma}_2}\left(\frac{1}{2}\lVert\dot{\Theta}_2(t)\rVert^2_F - \Gamma_\gamma\Gamma_\beta\mathcal{N}^{-1} L(\Theta_2(t))\right).
    \label{Eq32b}
\end{align}
\end{subequations}
The Lagrangian functions in \eqref{Eq32a}-\eqref{Eq32b} act as the basis of high-order tuners in this letter. Using a cost function $J(\Theta_j)$ as the integral of the functions for some time interval $t_\theta$, the Euler-Lagrangian equation of $\frac{d}{dt}\nabla_{\dot{\Theta}_j}\mathscr{L}(\cdot) = \nabla_{\Theta_j}\mathscr{L}(\cdot)$ and neglecting the time derivative of the normalization $\dot{\mathcal{N}}$, the high-order tuners yield in,
\begin{subequations}
\begin{align}
    \Ddot{\Theta}_1^\top(t) + \Gamma_\beta\mathcal{N}\dot{\Theta}_1^\top(t) &= -\Gamma_\gamma\Gamma_\beta\mathcal{N}\nabla_{\Theta_1}L(\Theta_1(t)), \label{Eq33a} \\
    \Ddot{\Theta}_2^\top(t) + \Gamma_\beta\dot{\Theta}_2^\top(t) &= -\Gamma_\gamma\Gamma_\beta\mathcal{N}^{-1}\nabla_{\Theta_2} L(\Theta_2(t)),\label{Eq33b}
\end{align}
\end{subequations}
or similarly for \eqref{Eq33a}, we can write in the following fashion using a new variable $\Xi_1$,
\begin{align}
    \dot{\Theta}_1^\top(t) &= -\Gamma_\beta\mathcal{N}(\Theta_1^\top(t) - \Xi_1^\top(t)). \label{Eq34}
\end{align}
\begin{remark}\label{rem4}
    The normalization $\mathcal{N}$ in \eqref{Eq33a}-\eqref{Eq33b} is required for stability proof with $\mu\geq 2(\gamma_m/\beta_m) \|\mathbf{B}_a^\top\hat{\mathbb{L}}_mP_a\|_F^2$ while $\Gamma_\beta$ and $\Gamma_\gamma$ shows the damping and the forcing term of the methods, respectively. Also, $\exists Q_a=Q_a^\top\succeq 2I_a$ that solves $\mathbf{A}_a^\top P_a + P_a\mathbf{A}_a = -Q_a$ and $P_a\mathbf{B}_a = \mathbf{C}_a^\top$
\end{remark}
\begin{theorem}\label{thm3}
    Using the followers $\mathcal{G}_m$ in \eqref{Eq17}, the leader $\ell$ in \eqref{Eq19}, the adaptive laws of $\dot{\Xi}_1=\dot{\Theta}$ in \eqref{Eq22} and \eqref{Eq34}, and the controller $\eqref{Eq16a}$ for the given $\mathcal{N},\mu,\Gamma_\beta,\Gamma_\gamma,Q_a$, in Remark~\ref{rem4} results in bounded solutions $\Bar{e}_a\in\mathcal{L}^\infty$, $\Bar{e}\in\mathcal{L}^\infty$, $\Tilde{\Xi}_1\coloneqq(\Xi_1-\Theta_1^\ast)\in\mathcal{L}^\infty$, $(\Theta_1-\Xi)\in\mathcal{L}^\infty$ for arbitrary initial conditions with $\lim_{t\to\infty}\Bar{e}_a=0$. Also, if $\eta,\dot{\eta}\in\mathcal{L}^\infty$, then $\lim_{t\to\infty}\dot{\Xi}_1=0$, $\lim_{t\to\infty}\dot{\Theta}_1=0$, and $\lim_{t\to\infty}\Theta_1-\Xi_1=0$.
\end{theorem}
\begin{proof}
    Let us introduce \eqref{Eq22} into different fashion as follows,
    \begin{align}
        \dot{\Xi}_1^\top(t) = -\sign(\mathbf{k}_p)\Gamma_\gamma\mathbb{L}_m^\top\Bar{e}(t)\Bar{\eta}^\top(t),
        \label{Eq35}
    \end{align}
    or $\dot{\Xi}_1=-\Gamma_\gamma\nabla L(\Tilde{\Xi}_1)$ given $L(\cdot)$ in \eqref{Eq30} and thanks to Lemma~\ref{lem1}. Recalling the state error $\Bar{e}_a$ in \eqref{Eq20a} by defining $\Tilde{\Theta}_1\coloneqq\Theta_1-\Theta_1^\ast$ and $\Tilde{\Xi}_1 = \Xi_1 - \Theta_1^\ast$, then using tuner in \eqref{Eq34}, the Lyapunov candidate $V(\Bar{e}_a,\Theta_1,\Xi_1)$ is chosen as,
    \begin{align}
        V &= \Bar{e}^\top_a(t)P\Bar{e}_a(t) + \tr\Bigl[(\Theta_1(t) -\Xi_1(t))\Gamma_\gamma^{-1}(\Theta_1(t)-\Xi_1(t))^\top\Bigr] + \tr\Bigl[\Tilde{\Xi}_1(t)\Gamma_\gamma^{-1}\Tilde{\Xi}_1^\top(t)\Bigr], \label{Eq36}
    \end{align}
    having the time derivative of \eqref{Eq36} along the trajectory of \eqref{Eq20a}-\eqref{Eq20c} as,
    \begin{subequations}
    \begin{align}
        \dot{V} &= \dot{\Bar{e}}_a^\top P_a\Bar{e}_a + \Bar{e}_a^\top P_a\dot{\Bar{e}}_a + 2\tr\Bigl[(\dot{\Theta}_1 - \dot{\Xi}_1)\Gamma_\gamma^{-1}(\Theta_1-\Xi_1)^\top\Bigr] + 2\tr\Bigl[\dot{\Tilde{\Xi}}_1\Gamma_\gamma^{-1}\Tilde{\Xi}_1^\top\Bigr] \\
        &= \Bar{e}_a^\top\left[\mathbf{A}_a^\top P_a + P_a\mathbf{A}_a\right]\Bar{e}_a + 2\Bar{e}_a^\top P_a\hat{\mathbb{L}}_m \mathbf{B}_a\Tilde{\Theta}_1^\top\Bar{\eta} \nonumber\\
        &\quad+ 2\tr\left[\dot{\Theta}_1\Gamma_\gamma^{-1}(\Theta_1-\Xi_1)^\top\right] - 2\tr\left[\dot{\Xi}_1\Gamma_\gamma^{-1}(\Theta_1-\Xi_1)^\top\right] + 2\tr\left[\dot{\Xi}_1\Gamma_\gamma^{-1}\Tilde{\Xi}_1^\top\right] \\
        &= \Bar{e}_a^\top\left[\mathbf{A}_a^\top P_a + P_a\mathbf{A}_a\right]\Bar{e}_a + 2\Bar{e}_a^\top P_a\hat{\mathbb{L}}_m \mathbf{B}_a(\Theta_1-\Xi_1+\Tilde{\Xi}_1)^\top\Bar{\eta} \nonumber \\
        &\quad -2\tr\left[(\Theta_1-\Xi_1)\mathcal{N}\Gamma_\beta\Gamma_\gamma^{-1}(\Theta_1-\Xi_1)^\top\right] - 2\tr\left[\dot{\Xi}_1\Gamma_\gamma^{-1}(\Theta_1-\Xi_1)^\top\right] + 2\tr\left[\dot{\Xi}_1\Gamma_\gamma^{-1}\Tilde{\Xi}_1^\top\right] \\
        &= \Bar{e}_a^\top\left[\mathbf{A}_a^\top P_a + P_a\mathbf{A}_a\right]\Bar{e}_a + 2\Bar{e}_a^\top P_a\hat{\mathbb{L}}_m \mathbf{B}_a(\Theta_1-\Xi_1)^\top\Bar{\eta} \nonumber\\
        &\quad -2\tr\left[(\Theta_1-\Xi_1)\mathcal{N}\Gamma_\beta\Gamma_\gamma^{-1}(\Theta_1-\Xi_1)^\top\right] - 2\tr\left[\dot{\Xi}_1\Gamma_\gamma^{-1}(\Theta_1-\Xi_1)^\top\right] \\
        &= \Bar{e}_a^\top\left[\mathbf{A}_a^\top P_a + P_a\mathbf{A}_a\right]\Bar{e}_a + 4\Bar{e}_a^\top P_a\hat{\mathbb{L}}_m \mathbf{B}_a(\Theta_1-\Xi_1)^\top\Bar{\eta} \nonumber\\
        &\quad -2\tr\left[(\Theta_1-\Xi_1)\Gamma_\beta\Gamma_\gamma^{-1}(\Theta_1-\Xi_1)^\top\right] - 2\Bar{\eta}^\top\Bar{\eta}\tr\left[(\Theta_1-\Xi_1)\Gamma_\beta\Gamma_\gamma^{-1}(\Theta_1-\Xi_1)^\top\right] \\
        &= \Bar{e}_a^\top\left[\mathbf{A}_a^\top P_a + P_a\mathbf{A}_a\right]\Bar{e}_a + 4\Bar{e}_a^\top P_a\hat{\mathbb{L}}_m \mathbf{B}_a(\Theta_1-\Xi_1)^\top\Bar{\eta} \nonumber\\
        &\quad -\frac{2\beta_m}{\gamma_m}\tr\left[(\Theta_1-\Xi_1)(\Theta_1-\Xi_1)^\top\right] - \frac{2\beta_m\Bar{\eta}^\top\Bar{\eta}}{\gamma_m}\tr\left[(\Theta_1-\Xi_1)(\Theta_1-\Xi_1)^\top\right] \\
        &\leq -2\|\Bar{e}_a\|^2 + 4\Bar{e}_a^\top P_a\hat{\mathbb{L}}_m \mathbf{B}_a(\Theta_1-\Xi_1)^\top\Bar{\eta} -\frac{2\beta_m}{\gamma_m}\|(\Theta_1 - \Xi_1)^\top\|_F^2 - \frac{2\beta_m}{\gamma_m}\|(\Theta_1 - \Xi_1)^\top\Bar{\eta}\|^2_F \\
        &\leq -\|\Bar{e}_a\|^2 -\frac{2\beta_m}{\gamma_m}\|(\Theta_1 - \Xi_1)^\top\|_F^2 + 4\|\Bar{e}_a\|\|\mathbf{B}_a^\top\hat{\mathbb{L}}_mP_a\|_F\|(\Theta_1-\Xi_1)^\top\Bar{\eta}\|_F \nonumber\\
        &\quad - 4\|\mathbf{B}_a^\top\hat{\mathbb{L}}_mP_a\|_F^2\|(\Theta_1 - \Xi_1)^\top\Bar{\eta}\|^2_F -\|\Bar{e}_a\|^2 \\
        &\leq -\|\Bar{e}_a\|^2 -\frac{2\beta_m}{\gamma_m}\|(\Theta_1 - \Xi_1)^\top\|_F^2 - \left(\|\Bar{e}_a\| - 2\|\mathbf{B}_a^\top\hat{\mathbb{L}}_m P_a\|_F\|(\Theta_1-\Xi_1)^\top\Bar{\eta}\|_F\right)^2 \leq 0,
    \end{align}
    \end{subequations}
    using Cauchy–Schwarz inequality $\|AB\|\leq\|A\|_F\|B\|_2$. It concludes that the boundedness in Theorem~\ref{thm3} holds.
\end{proof}
\begin{remark}\label{rem5}
    A complementary proof for \eqref{Eq33b} using $\dot{\Xi}_2=-\Gamma_\gamma\mathcal{N}^{-1}\nabla L(\Tilde{\Xi}_2)$, $\mathcal{N}^{-1} = N_m$ and modifying \eqref{Eq20a} to accommodate $\mathcal{N}$, with the same Lyapunov function as \eqref{Eq36}, denoted as $V_2(\Bar{e}_a,\Theta_2,\Xi_2)$, results in bounded solutions in which the time derivative of \eqref{Eq36} along the trajectory of \eqref{Eq20a}-\eqref{Eq20c} using $\Xi_2$, yields in $\dot{V}_2 \leq N_m\dot{V}\leq 0$. Also, given the tuners of $\Theta_1$ and $\Theta_2$ in \eqref{Eq33a} and \eqref{Eq33b}, the values of $\Gamma_\beta$ and $\Gamma_\gamma$ should be chosen larger as systems becoming far away $(q > 2)$ from the leader $\ell$.
\end{remark}

\begin{figure*}[t!]
    \centering
    \subfloat[\label{F4a} Block diagram]{\includegraphics[width=.55\linewidth]{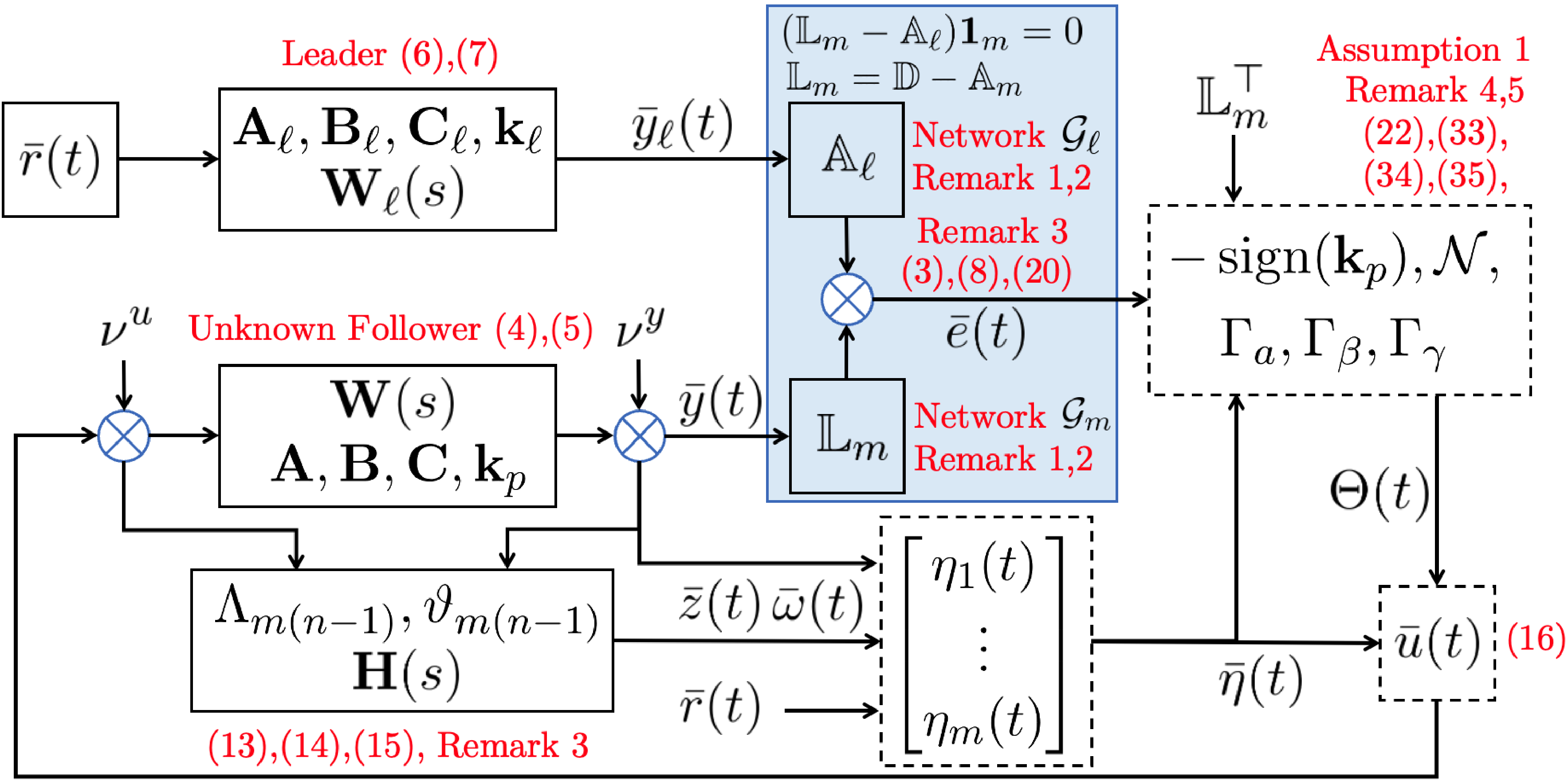}}\qquad
    \subfloat[\label{F4b} Algorithmic process]{\includegraphics[width=.2\linewidth]{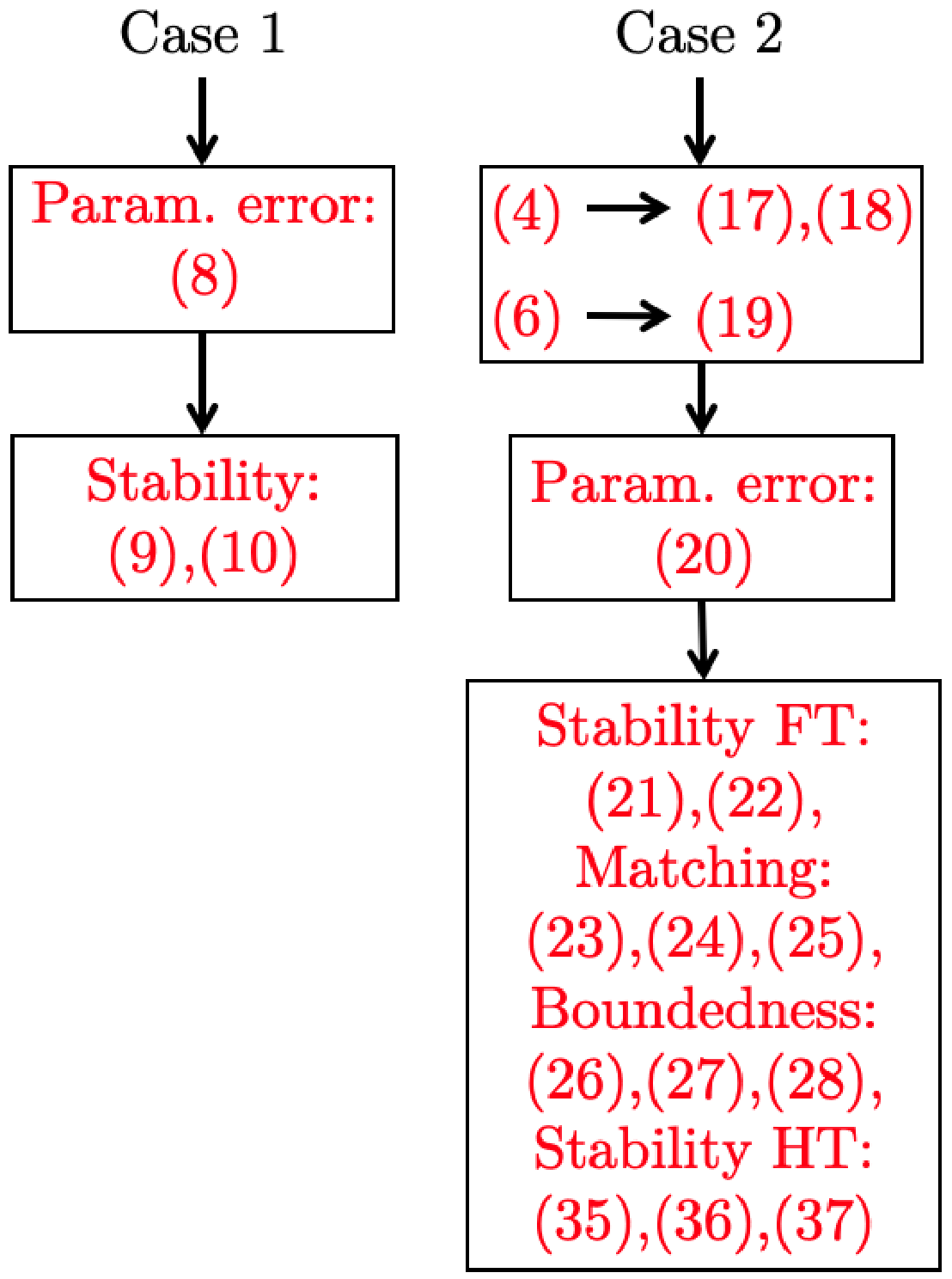}}
    \caption{(a) Scheme proposed in the paper where red color shows the equations (numbers inside brackets), remark, assumption, lemma or theorem associated to the blocks; and (b) The process of constructing the ideas in the paper.}
    \label{FG}
\end{figure*}

To end this section, we deliver the schemes graphically based on the equations provided, along with the algorithmic process, and give the reader the idea of implementing the proposed methods. The complete block diagram is portrayed in Figure \ref{FG}, such that

\section{Numerical Simulations and Findings}\label{S5}
\begin{figure*}[h!]
    \centering
    \includegraphics[width=.95\linewidth]{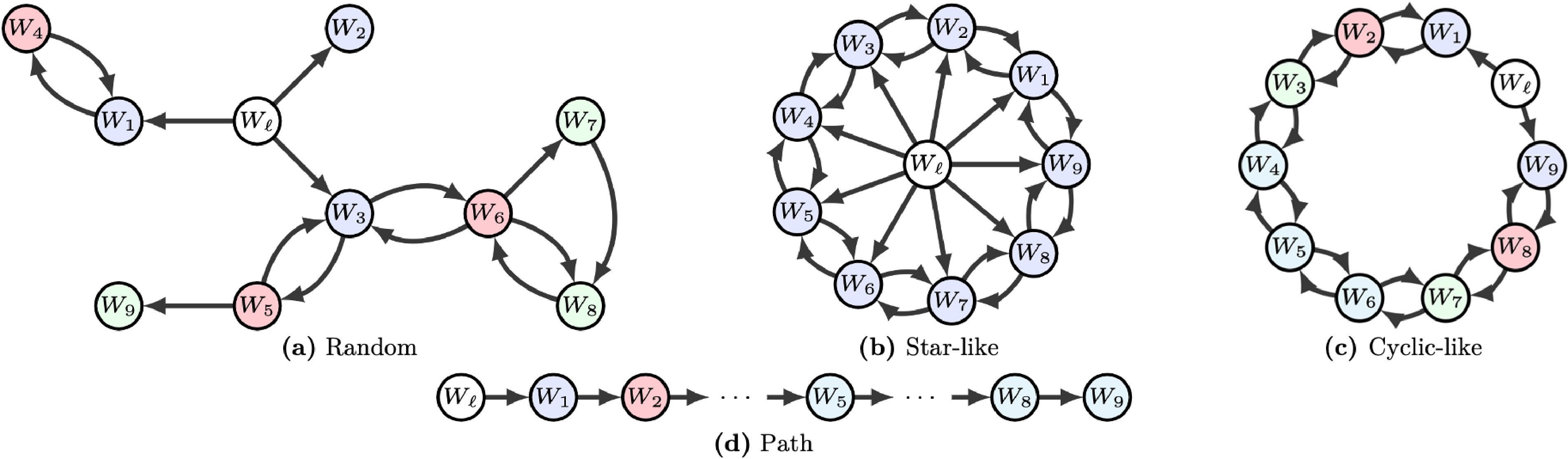}
    \caption{Four connected (not strongly connected) balanced ($\mathbb{D}=I_m$) networks used in the simulations.}
    \label{F3}
\end{figure*}

In this section, we simulate four different networks as in Figure~\ref{F1} and Figure~\ref{F3}, namely: Random, Star-like, Cyclic-like, and Path, with $m$ numbers of agents $m\in\{1,3,5,7,9,11,13\}$. The leader $\ell$ has a transfer function $W_\ell(s)$ and meanwhile, the $m$ followers are characterized by individual unstable transfer functions, each $W_i(s)$ defined as follows:
\begin{equation*}
    W_\ell(s) \coloneqq \frac{3s+3}{s^2+5s+6},\quad {W}_i(s)\coloneqq\frac{s+k+4}{(s-1-k)(s-2-k)}, \label{Eq38}
\end{equation*}
where $i=1,2,\dots,m$, when $m = 1$, $k=9i$ is used; when $m=3$, $k=4i-3$ is used; when $m=5$, $k=2i-1$ is used; when $m=7$, $k=(4i-1)/3$ is used; when $m=9$, $k=i$ is used; when $m=11$, $k=(4i+1)/5$ is used; and when $m=13$, $k=(2i+1)/3$ is used. The networks are balanced such that $(\mathbb{L}_m-\mathbb{A}_\ell)\mathbf{1}_m=0$ and $\mathbb{D} = I_m$. We design the weights $w_{ij}$ for the incoming measurements of node $i$ from its neighbors $j$ based on the level of $q$ where agents in $q\coloneqq 1$ gain more weights than those of $q=\{2,3,\dots\}$. The disturbances are $\nu^\alpha=[5,0.5]^\top\otimes\mathbf{1}_m$ for $\alpha=\{u,y\}$,$\forall t$. 

\begin{figure}[t!]
    \centering
    \subfloat[\label{F2a}$L_2-$norm and $L_\infty-$norm]{\includegraphics[width=.41\linewidth]{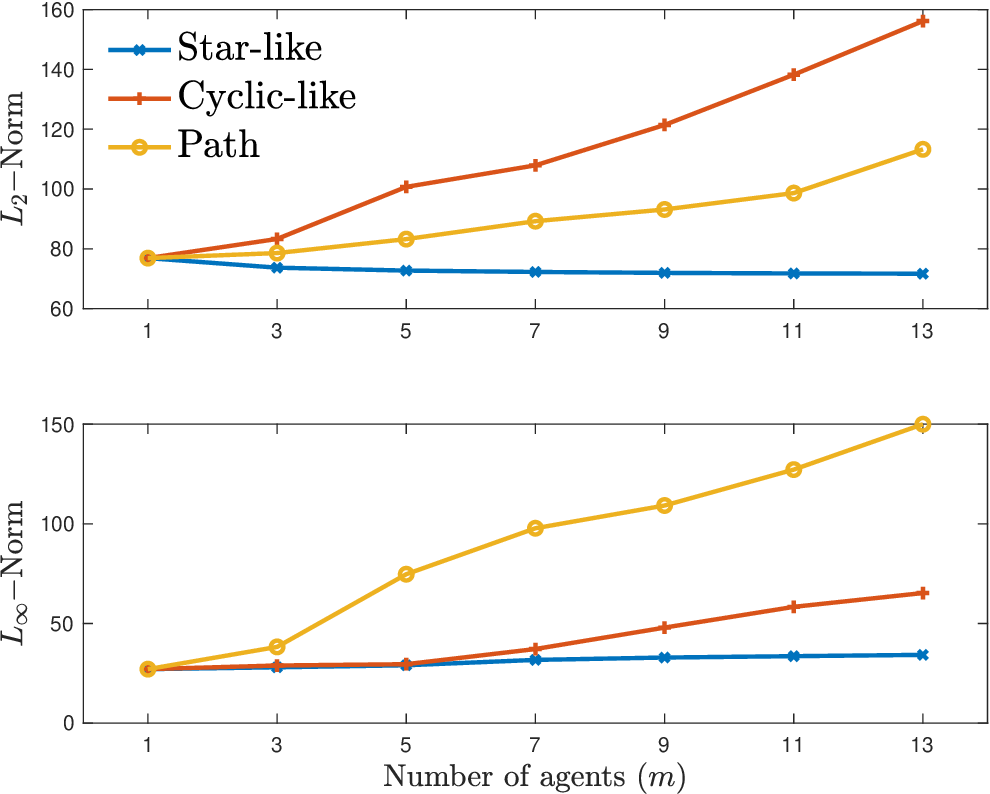}}\quad
    \subfloat[\label{F2b} $m= 9$]{\includegraphics[width=.4\linewidth]{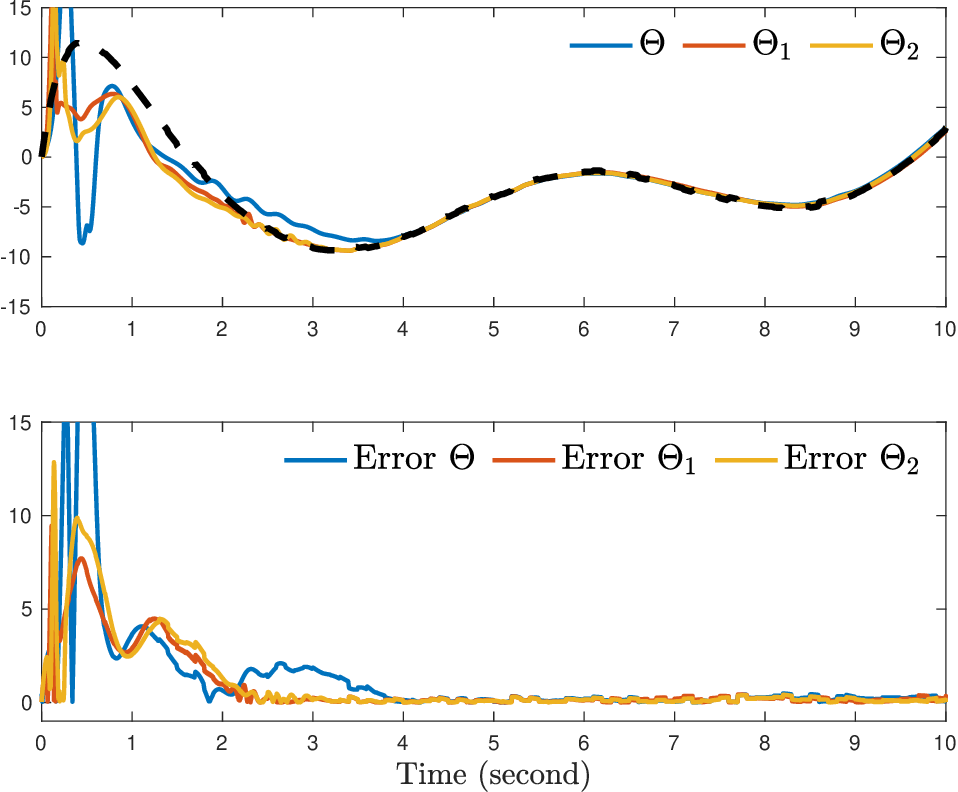}}
    \caption{(a) Performance measures of various $m$ in Figure~\ref{F3}; and (b) Random graph given by Figure~\ref{F1} with various tuners.}
    \label{F2}
\end{figure}

\begin{figure*}[t!]
    \centering
    \subfloat[\label{F4a} Star-like]{\includegraphics[width=.23\linewidth]{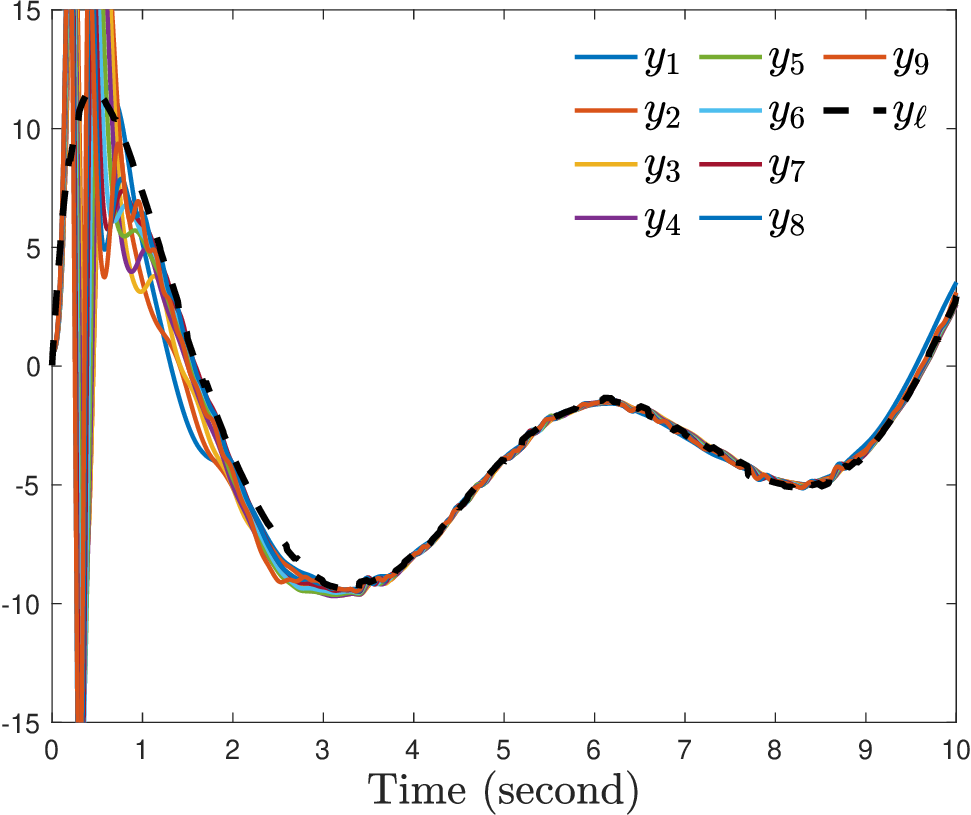}}\quad
    \subfloat[\label{F4b} Cyclic-like]{\includegraphics[width=.23\linewidth]{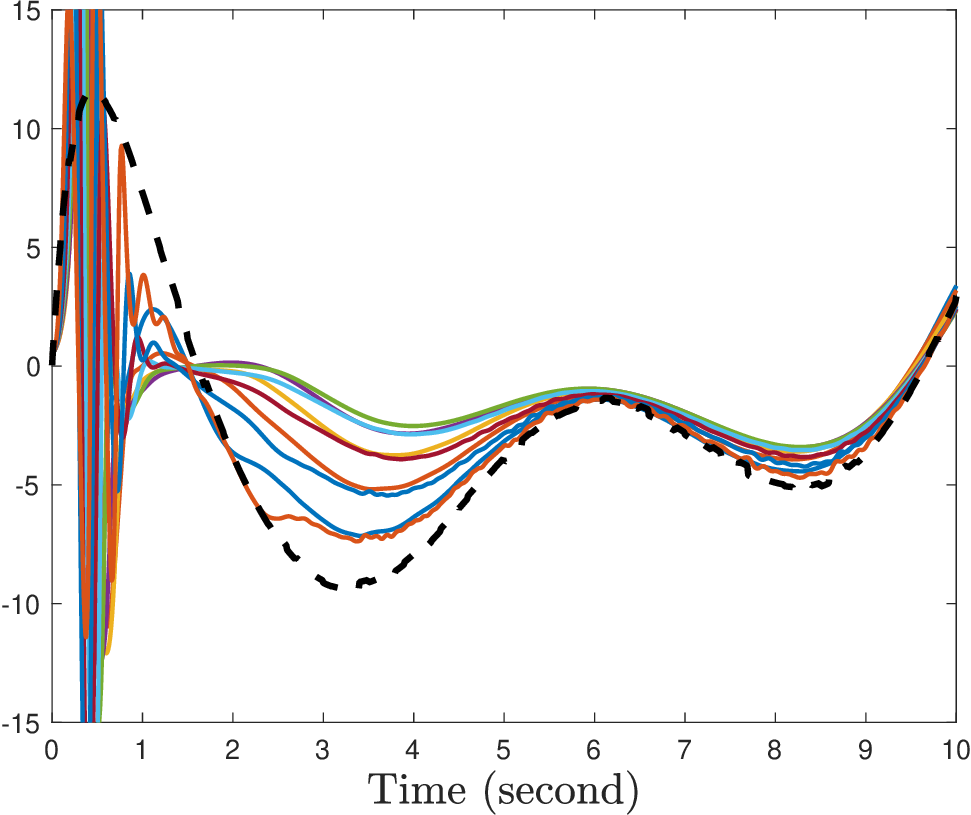}}\quad
    \subfloat[\label{F4c} Path]{\includegraphics[width=.23\linewidth]{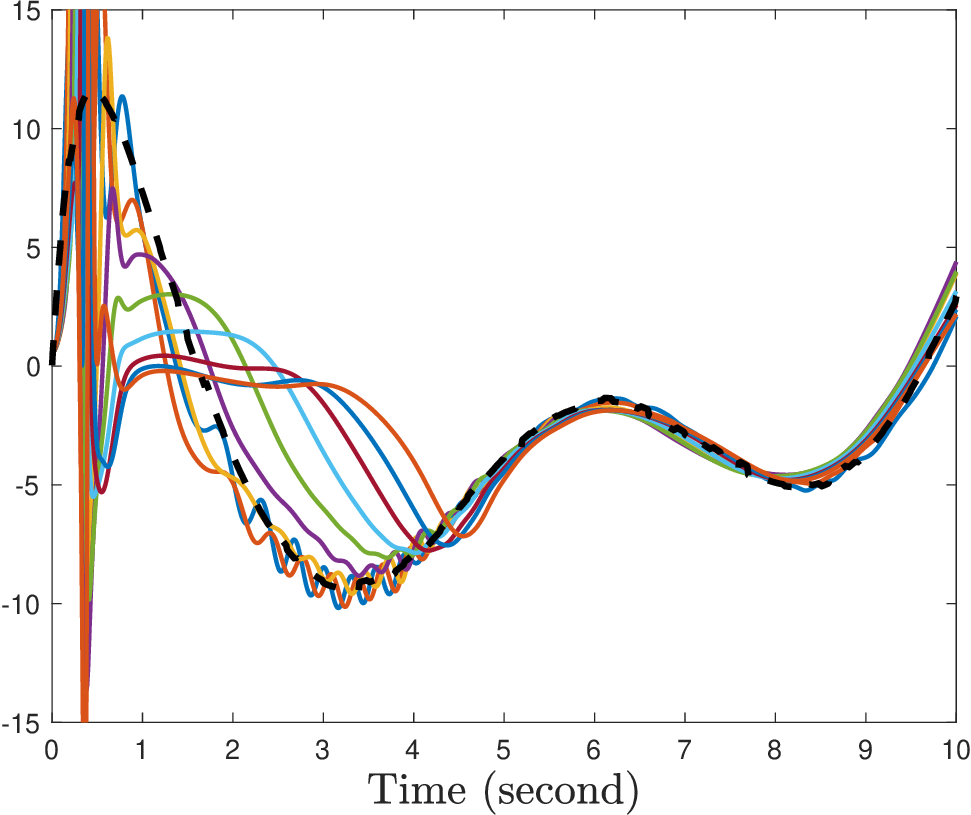}}\quad
    \subfloat[\label{F4d} Random]{\includegraphics[width=.23\linewidth]{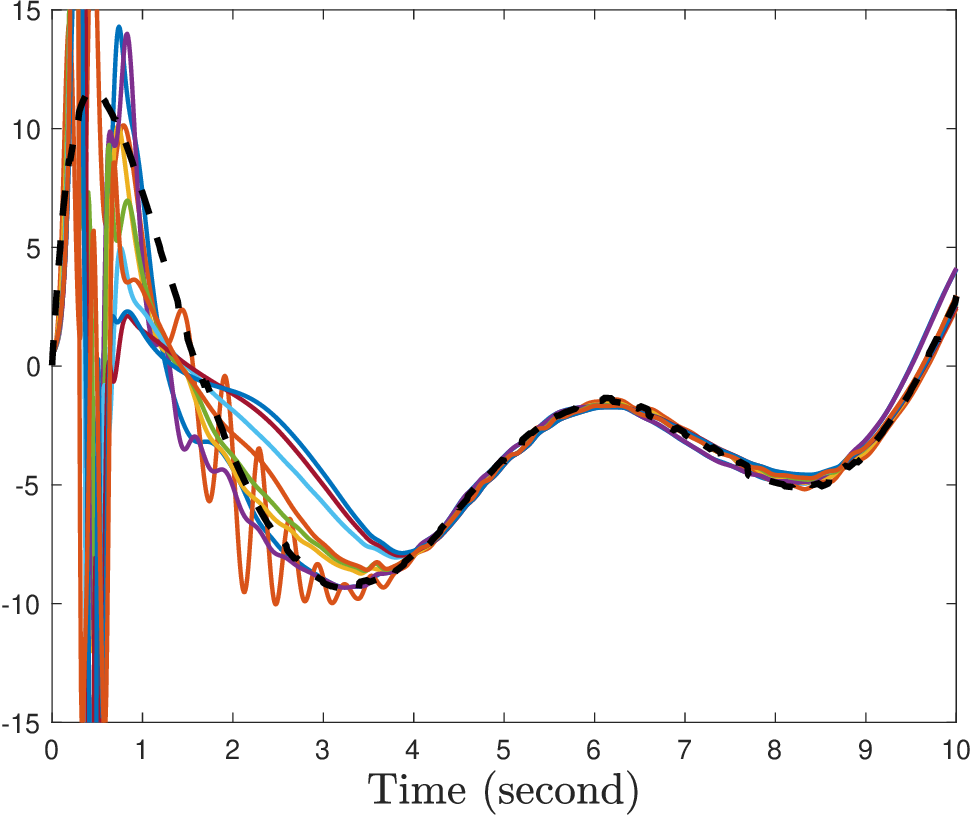}}
    \caption{Simulation results from network topologies with $m= 9$.}
    \label{F4}
\end{figure*}

For $m= 1$, this is the classic adaptive problem with weights of $\mathbb{L}_m=\mathbb{A}_\ell=I_m$ while for other $m$, we simulate three networks (Star-like, Cyclic-like, and Path) because they are comparable and the topologies are unchanged for various $m$. Furthermore, we compete three methods of tuners; first-order tuner $\Theta$ in \eqref{Eq22}, high-order tuners $\Theta_1$ in \eqref{Eq33a} and $\Theta_2$ in \eqref{Eq33b}. Keep in mind, due to Remark~\ref{rem5}, we adjust the constants of $\Gamma_\beta$ and $\Gamma_\gamma$ to be higher as $W_i$ becomes worse for the three tuners. Finally, we discuss the results using $L_2-$norm and $L_\infty-$norm of,
\begin{align*}
    \lVert f \rVert^2_2 = \int_{\mathbb{T}} |f(t)|^2 \, dt, \quad  \lVert f \rVert_\infty = \sup_{t \in \mathbb{T}} |f(t)|,
\end{align*}
in three different parts; ($\mathbf{a}$) Figure~\ref{F4} for $m= 9$; ($\mathbf{b}$) Figure~\ref{F2a} for various $m$; and ($\mathbf{c}$) Figure~\ref{F2b} and Table.~\ref{T2} for three tuners.

Part ($\mathbf{a}$). Among the networks, Star is the best since the whole agents connect directly to the leader with $w_{i\ell}=0.5,\forall i$ and $w_{ij}=0.25,\forall j\neq \ell$ weights. Regarding random-graph, the performance lies in the closeness to the leader (level of $q$) and the weights, with the most outer agents as the worst. It is confirmed for cyclic-graph that $W_4$, $W_5$, and $W_6$ are the worst and the latest to reach the consensus with the largest $q$. Furthermore, Path-graph has the highest $L_\infty-$norm since the consensus should wait for the preceding agents to be the same as leader $W_\ell$. However, the highest $L_\infty-$norm happens for some initial time, and does not guarantee the the overall performance, ensured by the better $L_2-$norm for various $m$.

Part ($\mathbf{b}$). For $m= 1$, any networks define the classic MRAC problem, resulting the same errors. For the Star, it is interesting as $m$ is increasing, it yields the smaller $L_2-$norm because the systems in between ($W_1,W_9$) are becoming less apart which do not happen to the other networks. Regarding Path, even though for some initial time the $L_\infty-$norm for Path is more than that of the Cyclic, the $L_2-$norm for various $m$ of Path outperforms its counterpart. This is due to the fact that the measurement gained for $W_i$ is $w_{ij}=1$ from the starting $W_\ell$ while for the Cyclic, the consensus is slightly slower due to many communications from non-leader agents.

Part ($\mathbf{c}$). The advantages of high-order tuners (HT) lies in 1) the stability with time-varying regressors and 2) an accelerated method with $\mathcal{O}(1/\sqrt{\epsilon})$ for a convex loss function, as opposed to the classic gradient descent $\mathcal{O}(1/\epsilon)$ \cite{R20}. The result shows that the modified high-order tuners, $\Theta_1$ and $\Theta_2$, perform better than the standard gradient-based $\Theta$. Note that, for Part ($\mathbf{a}$) and ($\mathbf{b}$), we use $\Theta$ in \eqref{Eq22}.
\begin{table}[t!]
    \centering
    \caption{$L_2-$norm and $L_\infty-$norm of the random graph.}
    \label{T2}
    \begin{adjustbox}{width=.5\textwidth}
    \begin{tabular}{@{}lccc@{\hspace{0.5cm}}ccc@{}} 
    \toprule
         & \multicolumn{3}{c}{$L_2$-Norm} & \multicolumn{3}{c}{$L_\infty$-Norm} \\
         \cmidrule(r){2-4} \cmidrule(l){5-7}
        & $\Theta$ & $\Theta_1$ & $\Theta_2$ & $\Theta$ & $\Theta_1$ & $\Theta_2$ \\
    \midrule
    Random Graph & 104.76 & 97.72 & 98.26 & 58.45 & 37.61 & 39.79 \\
    \bottomrule
    \end{tabular}
    \end{adjustbox}
\end{table}
\begin{table}[t!]
    \centering
    \caption{$L_2-$ norm and $L_\infty-$norm from Figure~\ref{F3} graphs and various numbers of $m$.}
    \label{T3}
    \begin{adjustbox}{width=.95\textwidth}
    \begin{tabular}{c|ccccccc|ccccccc}
        \toprule
         & \multicolumn{7}{c}{$L_2-$Norm for different values of $m$} & \multicolumn{7}{c}{$L_\infty-$Norm for different values of $m$} \\
          & $1$ & $3$ & $5$ & $7$ & $9$ & $11$ & $13$ & $1$ & $3$ & $5$ & $7$ & $9$ & $11$ & $13$ \\ \midrule
         Star & 76.88 & 73.68 & 72.71 & 72.25 & 71.94 & 71.76 & 71.66 & 27.10 & 28.03 & 29.04 & 31.74 & 32.93 & 33.61 & 34.27 \\
         Cyclic & 76.88 & 83.29 & 100.71 & 107.91 & 121.42 & 138.23 & 156.17 & 27.10 & 28.93 & 29.59 & 37.20 & 47.95 & 58.36 & 65.30 \\
         Path & 76.88 & 78.57 & 83.24 & 89.23 & 93.15 & 98.64 & 113.27 & 27.10 & 38.26 & 74.73 & 97.80 & 109.21 & 127.23 & 149.99\\
        \bottomrule
    \end{tabular}
    \end{adjustbox}
\end{table}

\section{Conclusion and Future Work}\label{S6}
We study the distributed adaptive control with network perspective using different agents and tuners. We provide the mathematical foundation, the designs, and the comparative illustrations. The results conclude some interesting trends based on the topologies and the increasing agents. There exists a stable network in Star-like while the highest $L_\infty-$norm of Path does not reflect the overall performance, outperforming Cyclic-like with lower $L_2-$norm. Moreover, we also show that the modified high-order tuners outperform the gradient-based method. Finally, the future research focuses on adding the delays or packet loss in communication networks and using the control-oriented learning \cite{R21}.

\section*{Acknowledgments}
The authors would like to thank Professor Anuradha M. Annaswamy and Peter A. Fisher from MIT for her valuable comments and insights from her course on Adaptive Control and its connection with Machine Learning.

\bibliographystyle{unsrt}  
\bibliography{EN-Bib}  

\newpage
\appendix
\section{Explanation of Lemma~\ref{lem2}}\label{A1}
Here, we expand in detail how to obtain \eqref{Eq23} such that $\Bar{y}\equiv\Bar{y}_\ell$ using the transfer function of \eqref{Eq4} and \eqref{Eq6}, which are denoted as $\mathbf{W}=\mathbf{k}_p\mathbf{N}\mathbf{D}^{-1}$ and $\mathbf{W}_\ell=\mathbf{k}_\ell\mathbf{N}_\ell\mathbf{D}_\ell^{-1}$, and the optimal control $\Bar{u}^\ast$, implying $\mathbf{k}^\ast$, $\Psi_d^{\ast}$, $\Phi_d^{\ast}$, and $T_d^{\ast}$. Also we define $\mathbf{H}\coloneqq\mathbf{N}_\Lambda\mathbf{D}_{\Lambda}^{-1}$ with $\mathbf{N}^\top_\Lambda(s) = \diag\{n_{\lambda_1}^\top(s),\dots,n_{\lambda_{m}}^\top(s)\}$ and $\mathbf{D}_\Lambda(s) = \diag\{d_{\lambda_1}(s),\dots,d_{\lambda_{m}}(s)\}$. The derivation is conducted as follows,
\begin{align}
    \mathbf{W}\left(\mathbf{k}^\ast\Bar{r} + \Psi_d^{\ast\top}\Bar{z} + \Phi_d^{\ast\top}\Bar{\omega} + T_d^\ast\Bar{y}\right) &= \mathbf{W}_\ell\Bar{r} \tag{A1}\label{EqA1}\\
    \mathbf{W}\left(\Psi_d^{\ast\top}\Bar{z} + \Phi_d^{\ast\top}\Bar{\omega} + T_d^\ast\Bar{y}\right) &= \left(\mathbf{W}_\ell - \mathbf{W}\mathbf{k}^\ast\right)\Bar{r} \tag{A2}\label{EqA2}
\end{align}
and by recalling \eqref{Eq11}, \eqref{Eq12} and $\Bar{r}\coloneqq\mathbf{W}_\ell^{-1}\mathbf{W}\Bar{u}^\ast$, we have
\begin{align}
    \mathbf{W}\left(\Psi_d^{\ast\top}\mathbf{H}\Bar{u}^\ast + \Phi_d^{\ast\top}\mathbf{H}\mathbf{W}\Bar{u}^\ast + T_d^\ast\mathbf{W}\Bar{u}^\ast\right) &= \left(\mathbf{W}_\ell - \mathbf{W}\mathbf{k}^\ast\right)\mathbf{W}_\ell^{-1}\mathbf{W}\Bar{u}^\ast \tag{A3}\label{EqA3}\\
    \mathbf{W} \left(\Psi_d^{\ast\top}\mathbf{N}_\Lambda\mathbf{D}_\Lambda^{-1}\Bar{u}^\ast + \Phi_d^{\ast\top}\mathbf{N}_\Lambda\mathbf{D}_\Lambda^{-1}\mathbf{k}_p\mathbf{N}\mathbf{D}^{-1}\Bar{u}^\ast + T_d^\ast\mathbf{k}_p\mathbf{N}\mathbf{D}^{-1}\Bar{u}^\ast\right) &= \mathbf{W}\Bar{u}^\ast  - \mathbf{W}\mathbf{k}^\ast\mathbf{W}_\ell^{-1}\mathbf{W}\Bar{u}^\ast \tag{A4}\label{EqA4}\\
    \Psi_d^{\ast\top}\mathbf{N}_\Lambda\mathbf{D}_\Lambda^{-1} + \Phi_d^{\ast\top}\mathbf{N}_\Lambda\mathbf{D}_\Lambda^{-1}\mathbf{k}_p\mathbf{N}\mathbf{D}^{-1} + T_d^\ast\mathbf{k}_p\mathbf{N}\mathbf{D}^{-1} &= I_m - \mathbf{k}^\ast\mathbf{W}_\ell^{-1}\mathbf{W}\tag{A5}\label{EqA5}
\end{align}
in which after multiplying both sides with $\mathbf{D}_\Lambda\mathbf{D}$, we have
\begin{align}
    \Psi_d^{\ast\top}\mathbf{N}_\Lambda\mathbf{D} + \Phi_d^{\ast\top}\mathbf{N}_\Lambda\mathbf{k}_p\mathbf{N} + T_d^\ast\mathbf{D}_\Lambda\mathbf{k}_p\mathbf{N} &= \mathbf{D}_\Lambda\left(\mathbf{D} - \mathbf{k}^\ast\mathbf{W}_\ell^{-1}\mathbf{k}_p\mathbf{N}\right) \tag{A6}\label{EqA6}\\
    \Psi_d^{\ast\top}\mathbf{N}_\Lambda\mathbf{D} + \left(\Phi_d^{\ast\top}\mathbf{N}_\Lambda + T_d^\ast\mathbf{D}_\Lambda\right)\mathbf{k}_p\mathbf{N} &= \mathbf{D}_\Lambda\left[\mathbf{D} - \mathbf{k}^\ast\left(\mathbf{k}_\ell\mathbf{N}_\ell\mathbf{D}_\ell^{-1}\right)^{-1}\mathbf{k}_p\mathbf{N}\right]\tag{A7}\label{EqA7}
\end{align}

\section{Proof of Lemma~\ref{lem2}}\label{A2}
Here, we continue the last part in which if $\mathbf{N}(s)$ and $\mathbf{D}(s)$ are coprime, then the rank of $\mathbb{S}$, defined as $\rho(\mathbb{S})$, be full rank and \eqref{Eq25} has a solution, otherwise the solution is not unique. However, given $\mathbf{N}(s)=\Bar{\mathbf{N}}(s)\mathbf{P}(s)$ and $\mathbf{D}(s)=\Bar{\mathbf{D}}(s)\mathbf{P}(s)$, for some polynomial $\mathbf{P}(s)=\diag\{p_1(s),\dots,p_m(s)\}$ of degree $r_p>0$, such that $\Bar{\mathbf{N}}(s)$ and $\Bar{\mathbf{D}}(s)$ are coprime, then $\eqref{Eq23}$ becomes,
\begin{align}\begin{aligned}
    \Psi_d^{\ast\top}\mathbf{N}_\Lambda\Bar{\mathbf{D}} + \left(\Phi_d^{\ast\top}\mathbf{N}_\Lambda + T_d^\ast\mathbf{D}_\Lambda\right)\mathbf{k}_p\Bar{\mathbf{N}} &= \mathbf{D}_\Lambda\left[\Bar{\mathbf{D}} - \mathbf{k}^\ast\left(\mathbf{k}_\ell\mathbf{N}_\ell\mathbf{D}_\ell^{-1}\right)^{-1}\mathbf{k}_p\Bar{\mathbf{N}}\right]
    \end{aligned} \tag{B1}\label{EqB1}
\end{align}
from which this new equation is satisfied $\Bar{\mathbb{S}}_n\Bar{\Theta}_s^\ast = \Bar{\Pi}_n$ where $\Bar{\mathbb{S}}_n\in\mathbb{R}^{(r_m-r_p)\times r_m}$ and $\Bar{\Pi}_n\in\mathbb{R}^{r_m-r_p}$. Due to the fact that $\Bar{\mathbf{N}}(s)$ and $\Bar{\mathbf{D}}(s)$ are coprime, then $\rho(\mathbb{S}_n)=r_m-r_p$, meaning $\Bar{\mathbb{S}}_n\Bar{\Theta}_s^\ast = \Bar{\Pi}_n$ has a solution which is not unique since $\Bar{n}>\Bar{n}-r_p$. However, with the help of, 
\begin{align}\begin{aligned}
    \mathbf{Q}_1&=\diag\{[I_{n_1-1},0_{(n_1-1)\times n_1}],\dots,[I_{n_m-1},0_{(n_m-1)\times n_m}]\} \\
    \mathbf{Q}_2&=\diag\{[0_{(n_1-1)\times n_1},I_{n_1-1}],\dots,[0_{(n_m-1)\times n_m},I_{n_m-1}]\} \\
    \mathbf{Q}_3&=\diag\{[0_{1\times(n_1-1)},1,0_{1\times(n_1-1)}],\dots,[0_{1\times(n_m-1)},1,0_{1\times(n_m-1)}]\} \end{aligned}
    \tag{B2}\label{EqB2}
\end{align}
such that $\mathbf{Q}_1\Bar{\Theta}_s\coloneqq \Psi^\ast$ and $\mathbf{Q}_2\Bar{\Theta}_s - \mathbf{Q}_3\Bar{\Theta}_s\Bar{\mathbf{D}}_\lambda \coloneqq \Phi^\ast$ in which $\Bar{\mathbf{D}}_\lambda$ is vector element of the Laplace coefficients starting from $n-2$, denoted as $\Bar{\mathbf{D}}_\lambda=\diag\{\Bar{d}_{\lambda_1},\dots,\Bar{d}_{\lambda_m}\}$, then there exist $\Theta^\ast$ such that \eqref{Eq25} is satisfied.

\section{Proof of \eqref{Eq27} and \eqref{Eq28} in Theorem~\ref{thm1}}\label{A3}
Operating both sides of $\eqref{Eq23}$ with $\mathbb{L}_m\Bar{y}$ and considering $\Bar{y} = \mathbf{k}_p\mathbf{N}(s)\mathbf{D}^{-1}(s)\Bar{u}$, $\Bar{y}_\ell = \mathbf{k}_\ell\mathbf{N}_\ell(s)\mathbf{D}_\ell^{-1}(s)\Bar{r}$, and $\mathbf{H}(s)\coloneqq\mathbf{N}_\Lambda(s)\mathbf{D}_{\Lambda}^{-1}(s)$, we have the following,
\begin{align}
    \Psi_d^{\ast\top}\mathbf{N}_\Lambda\mathbf{D}\mathbb{L}_m\Bar{y} + \left(\Phi_d^{\ast\top}\mathbf{N}_\Lambda + T_d^\ast\mathbf{D}_\Lambda\right)\mathbf{k}_p\mathbf{N}\mathbb{L}_m\Bar{y} &= \mathbf{D}_\Lambda\left(\mathbf{D}\mathbb{L}_m\Bar{y} - \mathbf{k}^\ast\mathbf{W}_\ell^{-1}\mathbf{k}_p\mathbf{N}\mathbb{L}_m\Bar{y}\right)\tag{C1}\label{EqC1} \\
    \Psi_d^{\ast\top}\mathbf{N}_\Lambda\mathbb{L}_m\mathbf{k}_p\mathbf{N}\Bar{u} + \left(\Phi_d^{\ast\top}\mathbf{N}_\Lambda + T_d^\ast\mathbf{D}_\Lambda\right)\mathbf{k}_p\mathbf{N}\mathbb{L}_m\Bar{y} &= \mathbf{D}_\Lambda\left(\mathbb{L}_m\mathbf{k}_p\mathbf{N}\Bar{u} - \mathbf{k}^\ast\mathbf{W}_\ell^{-1}\mathbf{k}_p\mathbf{N}\mathbb{L}_m\Bar{y}\right)\tag{C2}\label{EqC2} 
\end{align}
which can also be constructed as follows,
\begin{align}
    \mathbf{D}_\Lambda\mathbb{L}_m\mathbf{k}_p\mathbf{N}\Bar{u} &= \Psi_d^{\ast\top}\mathbf{N}_\Lambda\mathbb{L}_m\mathbf{k}_p\mathbf{N}\Bar{u} + \left(\Phi_d^{\ast\top}\mathbf{N}_\Lambda + T_d^\ast\mathbf{D}_\Lambda\right)\mathbf{k}_p\mathbf{N}\mathbb{L}_m\Bar{y} + \mathbf{D}_\Lambda\mathbf{k}^\ast\mathbf{W}_\ell^{-1}\mathbf{k}_p\mathbf{N}\mathbb{L}_m\Bar{y} \tag{C3}\label{EqC3} \\
    \mathbb{L}_m\Bar{u} &= \mathbb{L}_m\Psi_d^{\ast\top}\mathbf{H}\Bar{u} + \mathbb{L}_m\Phi_d^{\ast\top}\mathbf{H}\Bar{y} + \mathbb{L}_m T_d^\ast\Bar{y} + \mathbb{L}_m\mathbf{k}^\ast\mathbf{W}_\ell^{-1}\Bar{y} \tag{C4}\label{EqC4} 
\end{align}
then, we add this zero term $(\mathbb{L}_m-\mathbb{A}_\ell)\mathbf{k}^\ast\Bar{r}\coloneqq 0$ so that it does not change the equation, 
\begin{align}
    \mathbb{L}_m\Bar{u} &= \mathbb{L}_m\Psi_d^{\ast\top}\mathbf{H}\Bar{u} + \mathbb{L}_m\Phi_d^{\ast\top}\mathbf{H}\Bar{y} + \mathbb{L}_m T_d^\ast\Bar{y} + \mathbb{L}_m\mathbf{k}^\ast\mathbf{W}_\ell^{-1}\Bar{y} + (\mathbb{L}_m-\mathbb{A}_\ell)\mathbf{k}^\ast\Bar{r} \tag{C5}\label{EqC5} \\
    \mathbb{L}_m\Bar{u} &= \mathbb{L}_m\underbrace{\left(\Psi_d^{\ast\top}\mathbf{H}\Bar{u} + \Phi_d^{\ast\top}\mathbf{H}\Bar{y} + T_d^\ast\Bar{y} + \mathbf{k}^\ast\Bar{r}\right)}_{\Bar{u}} + \mathbb{L}_m\mathbf{k}^\ast\mathbf{W}_\ell^{-1}\Bar{y} -\mathbb{A}_\ell\mathbf{k}^\ast\Bar{r} \tag{C6}\label{EqC6} \\
    0 &= \mathbb{L}_m\mathbf{k}^\ast\mathbf{W}_\ell^{-1}\Bar{y} -\mathbb{A}_\ell\mathbf{k}^\ast\mathbf{W}_\ell^{-1}\Bar{y}_\ell \tag{C7}\label{EqC7} 
\end{align}


\end{document}